\documentclass[12pt]{article}
\usepackage{amsmath,amssymb,tikz-cd}
\usepackage{hyperref}
\usepackage{cite}
\usepackage{tikz}
\usetikzlibrary{arrows,chains,matrix,positioning,scopes,decorations,fadings}

\makeatletter
\tikzset{join/.code=\tikzset{after node path={%
\ifx\tikzchainprevious\pgfutil@empty\else(\tikzchainprevious)%
edge[every join]#1(\tikzchaincurrent)\fi}}}
\makeatother

\tikzset{>=stealth',every on chain/.append style={join},3
         every join/.style={->}}
\tikzset{
    >=stealth',
    punkt/.style={
           rectangle,
           rounded corners,
           draw=black, very thick,
           text width=6.5em,
           minimum height=2em,
           text centered},
    pil/.style={
           ->,
           thick,
           shorten <=2pt,
           shorten >=2pt,}
}

\newcommand{\bea}{\begin{eqnarray}}
\newcommand{\eea}{\end{eqnarray}}
\newcommand{\be}{\begin{equation}}
\newcommand{\ee}{\end{equation}}
\newcommand{\nn}{\nonumber}
\newcommand{\Tr}{\textrm{Tr}}

\newcommand{\dl}{\textrm{d}_L}
\newcommand{\dr}{\textrm{d}_R}
\def\A{{\mathcal{A}}}
\def\gh{{\rm gh}}
\def\g{{\mathfrak g}}
\def\kin{{\cal K}}
\def\gen{{\cal D}^\dagger}
\def\D{{\cal D}}
\def\so{{\mathfrak {so}}}
\def\Re{{\mathbb R}}
\def\Gerst{{\mathsf G}}

\def\E{{\cal E}}
\def\vol{{\rm vol}}
\def\Vect{{\rm Vect}}
\def\M{{\cal M}}
\def\C{{\mathbf C}}
\def\PHI{{\mathbf{\Phi}}}
\def\dgLa{{\cal L}} 
\def\cdga{{\cal C}} 

\DeclareMathAlphabet{\mathpzc}{OT1}{pzc}{m}{it}

\usepackage{amsthm}
\usepackage{amsmath}
\usepackage{amssymb}
\usepackage{amsfonts}

\theoremstyle{definition}

\usetikzlibrary{arrows}

\begin{document}
\begin{flushright}
UUITP-18/25 \\
MIT-CTP/5878
\end{flushright}

\begin{center} \Large
{\bf Off-shell double copy theories in BV}
 \\[12mm] \normalsize
{\bf Maor Ben-Shahar${}^{a}$, Francesco Bonechi${}^{b}$ and Maxim Zabzine${}^{c,d}$} \\
[8mm]
{\small\it ${}^a$  MIT Center for Theoretical Physics - a Leinweber Institute,\\
Cambridge, MA 02139, USA
\\}
{\small\it ${}^b$ INFN Sezione di Firenze, Via G. Sansone 1,\\
 50019 Sesto Fiorentino, Firenze, Italy \\}
 {\small\it ${}^c$Department of Physics and Astronomy, Uppsala University,\\ Box 516, SE-75120 Uppsala, Sweden\\}
{\small\it ${}^d$Centre for Geometry and Physics, Uppsala University,\\
 Box 480, SE-75106 Uppsala, Sweden\\}
\end{center}
\vspace{7mm}

\begin{abstract}
We present a construction of the double copy for gauge theories that exhibit off-shell color-kinematics duality within the Batalin-Vilkovitsky (BV) formalism. As illustrative examples, we consider the double copies of Chern-Simons theory, four-dimensional BF theory, and two-dimensional Yang-Mills theory, and we discuss possible gravity interpretations for these cases. We formalize the construction and demonstrate that Kodaira-Spencer gravity, K\"ahler gravity, and their generalizations, fit naturally within this framework. In particular, Kodaira-Spencer gravity emerges as a gauge theory describing deformations of generalized complex structures, while the double copy of Chern-Simons theory captures the deformation of the Courant bracket.
\end{abstract}

\eject

\tableofcontents

\section{Introduction}

The double-copy construction typically produces scattering amplitudes in theories of gravity from building blocks of gauge-theory scattering amplitudes \cite{Bern:2008qj,Bern:2010ue}. It goes beyond the Kawai-Lewellen-Tye (KLT) relations \cite{Kawai:1985xq} in that it applies to loops as well as to many theories of gravity, for a review see ref. \cite{Bern:2019prr}. The double copy takes as input theories that obey the color-kinematics duality, the condition that the kinematic numerators of cubic Feynman graphs obey the same identities as the color factors. In some cases, this duality is known to arise from a kinematic Lie algebra in terms of which the numerators can be written \cite{Monteiro:2011pc,Boels:2013bi,Cheung:2016prv,Cheung:2020djz,Ben-Shahar:2021zww,Cheung:2022mix,Ben-Shahar:2024dju}, implying the duality holds off-shell. 

The motivations for studying off-shell color-kinematics duality are multifold. First, the associated kinematic algebras are new, infinite-dimensional structures that are interesting in their own right. Second, off-shell color-kinematics duality allows for the construction of off-shell double copies, which may reveal hidden structures within the resulting gravity theories. In fact, in general the validity of the double-copy construction to all loop orders remains conjectural, but in cases where off-shell double copies can be constructed they directly produce actions for the double-copy theory implying that the double copy works to all loop orders \cite{Cheung:2020djz,Ben-Shahar:2021zww,Cheung:2022mix,Borsten:2023paw}. These off-shell constructions may allow to further extend the double copy to curved backgrounds \cite{Diwakar:2021juk,Herderschee:2022ntr,Cheung:2022pdk,Beetar:2024ptv} or establish a link with the double-copies of exact solutions on which there has been a large body of work \cite{Monteiro:2014cda,Luna:2015paa,Luna:2016hge,Bahjat-Abbas:2017htu,Carrillo-Gonzalez:2017iyj,Berman:2018hwd,CarrilloGonzalez:2019gof,Goldberger:2019xef,Huang:2019cja,Bahjat-Abbas:2020cyb,Easson:2020esh,Emond:2020lwi, Godazgar:2020zbv,Chacon:2021wbr,Chacon:2020fmr,Alfonsi:2020lub, Monteiro:2020plf, White:2020sfn, Elor:2020nqe,Pasarin:2020qoa, Adamo:2021dfg, Easson:2022zoh,Dempsey:2022sls,CarrilloGonzalez:2022ggn,Armstrong-Williams:2022apo,Kent:2025pvu}.

For Yang-Mills theory in general dimensions, off-shell color-kinematics remains illusive, although some progress has been made on Lagrangians that produce on-shell numerators that obey the duality  \cite{Bern:2010yg,Tolotti:2013caa,Ben-Shahar:2022ixa}. Even  without fully off-shell color-kinematics duality though there has been promising progress on constructions of double-copy actions through Batalin-Vilkovitsky quantization and homotopy-algebraic techniques \cite{ Borsten:2020xbt,Borsten:2020zgj,Borsten:2021zir,Borsten:2021hua,Borsten:2021gyl,Borsten:2022ouu,Borsten:2022vtg,Borsten:2023reb,Borsten:2023ned,Bonezzi:2022yuh,Bonezzi:2022bse,Bonezzi:2023lkx,Bonezzi:2023pox,Bonezzi:2023ciu,Escudero:2022zdz,Szabo:2023cmv,Diaz-Jaramillo:2025gxw,Bonezzi:2025anl}.

The study of color-kinematics duality in the context of homotopy algebras was initiated by Reiterer in ref. \cite{Reiterer:2019dys}, in which it was observed that the color-kinematics duality emerges form the requirement that the propagator-numerator is a second-order operator. Such second-order operators, in their relationship to the underlying geometric structures in the theory were subsequently identified in a number of examples \cite{Ben-Shahar:2021doh,Ben-Shahar:2021zww}, including BF theory and Yang-Mills theory in two dimensions \cite{Ben-Shahar:2024dju}. The choice of such a propagator numerator is related to a gauge choice, and indeed the kinematic algebra and color-kinematics duality are both gauge dependent properties \cite{Bonezzi:2023pox,Armstrong-Williams:2024icu}. In absence of a second-order operator, duality satisfying numerators can none-the-less be constructed
\cite{Bjerrum-Bohr:2010pnr,Mafra:2011kj,Fu:2012uy,Mafra:2015vca,Bjerrum-Bohr:2016axv,Du:2017kpo,Chen:2017bug,Fu:2018hpu,Edison:2020ehu,He:2021lro,Bridges:2021ebs, Ahmadiniaz:2021fey,Cheung:2021zvb,Ahmadiniaz:2021ayd,Brandhuber:2021bsf,Cheung:2021zvb,Lee:2015upy,Bridges:2019siz,Chen:2022nei,Chen:2024gkj,Chen:2023ekh,Chen:2019ywi,Chen:2021chy} at tree as well as loop level \cite{Bern:2010ue,Carrasco:2011mn,Bern:2012uf,Boels:2013bi,Bjerrum-Bohr:2013iza,Bern:2013yya,Nohle:2013bfa,Mogull:2015adi,Mafra:2015mja,He:2015wgf,Johansson:2017bfl,Hohenegger:2017kqy,Mafra:2017ioj,Faller:2018vdz,Kalin:2018thp,Duhr:2019ywc,Geyer:2019hnn,Edison:2020uzf,Casali:2020knc,DHoker:2020prr,Carrasco:2020ywq,Bridges:2021ebs,Edison:2023ulf,Boels:2012ew,Yang:2016ear,Boels:2017ftb,Lin:2020dyj,Lin:2021qol,Lin:2021pne,Lin:2021lqo}, enabling the computation of amplitudes in many theories of gravity \cite{Broedel:2012rc, Chiodaroli:2013upa,Johansson:2014zca,Chiodaroli:2014xia,Johansson:2015oia,Chiodaroli:2015rdg,Johansson:2017srf,Ben-Shahar:2018uie,Chiodaroli:2017ngp,Chiodaroli:2018dbu,Johansson:2018ues,Azevedo:2018dgo,Johansson:2019dnu,Bautista:2019evw,Plefka:2019wyg,Pavao:2022kog, Mazloumi:2022nvi,Johansson:2025grx}.

 In this paper we utilize the kinematic algebras in BF and YM theories \cite{Ben-Shahar:2024dju} to construct novel double copy actions and discuss their connections to deformation of the Courant bracket.  In addition, we show that a generalized version of the double copy, in which the spacetime dimension is doubled, allows the construction of Kodaira-Spencer theory, K\"ahler gravity, and their generalizations, which can be defined on curved manifolds. These actions generalize the constructions seen for the double copy of Chern-Simons theory \cite{Ben-Shahar:2021zww,Bonezzi:2024dlv}, 
with the key ingredients in the BV action being a differential $D$, a second-order 
operator $\gen$ obeying $\gen D+D\gen = 0$ and $\gen{}^2=0$. Since $\gen$ is second order it gives rise to a Lie-algebra bracket $\{\A,\A\} = \gen(\A^2) - \gen\A \A - \A \gen \A $. If we restrict this bracket to ${\rm Im}(\gen)$ then on this Lie subalgebra we can define an invariant pairing $\langle \ \ \rangle$ compatible with these operators. 
The action is then
\begin{equation}
    S = \frac{1}{2}\langle \A, D \A\rangle + \frac{1}{6}\langle \A,\{\A,\A\}\rangle \ ,
\end{equation}
which resembles a Chern-Simons action and satisfies the master equation with symplectic form $\omega = \langle \delta\A , \delta \A\rangle$.
It is important for the construction that we introduce two more operators $\D$ and $D^\dagger$, obeying similar identities to above as well as $\D \gen+\gen \D=\Box=DD^\dagger +D^\dagger D$, altogether essentially mimicking the K\"ahler identities. Then after gauge fixing $\A\in \textrm{Im}(D^\dagger)$, two BRST symmetries emerge in our action,  see appendix \ref{a:BRST} where we also discuss an SL(2) freedom in rotating these BRST symmetries.
Generically, as observed in ref. \cite{Ben-Shahar:2021zww}, the off-shell double copy actions appear to be non-local, however, in cases where 2d YM or 4d BF theories are involved in the double copy, a local action can be found. Interestingly, for 4d BF theory we find a one-parameter family of kinematic algebras, giving some additional freedom in defining the double copy action.

The paper is organized as follows: 
In Section \ref{s:into-kinematic} we review the results form 
\cite{Ben-Shahar:2021zww} and we set the notion, following \cite{Ben-Shahar:2024dju}. 
In Section \ref{s:BV-double-CS} we offer the BV formulation of the double copy of Chern-Simons theory and explain that the proposal from \cite{Ben-Shahar:2021zww} is the gauge fixed version of this BV construction. Moreover, we suggest this double copy theory is related to gravity, in particular to deformations of a flat metric. 
In Section \ref{s:formal-CS} we generalize this construction to a formal Chern-Simons theory and to the double copy of it in flat space. As an illustration of the formal construction we present the example of 4D BF theory in Section \ref{s:BF}, which admits different kinematic algebras and correspondingly different realizations of a double copy. In Section \ref{s:2DYM} we briefly discuss the double copy of 2D Yang-Mills theory.
In Section \ref{s:formal} we formalize the construction of double copy theories within the BV formalism. The BV formulation of Kodaira-Spencer and K\"ahler gravity fit this framework as we explain in Section \ref{s:KS}. We also consider the generalization of these to other dimensions. 
Section \ref{s:end} summarizes the paper and discusses the open questions. 

\section{Kinematic Lie algebra and double copy}\label{s:into-kinematic}

Chern-Simons theory obeys the color-kinematics duality \cite{Ben-Shahar:2021zww}, and therefore can be used in the double-copy construction. We will review these results here and use them to introduce the notation we use throughout this paper.
The on-shell description of color-kinematics duality requires organizing Feynman diagrams in the gauge theory of interest into sums over cubic graphs,
\begin{equation}
    \mathcal{M} = \sum_{i\in\Gamma_3}\int_{l_i} \frac{n_i c_i}{D_i} \ ,
\end{equation}
where each graph is assigned a denominator, $D_i$, a color factor $c_i$ and a numerator $n_i$. The denominators are the propagators attached to the internal lines of the Feynman diagram, and are simply products of squared momenta $\sim p^2$. The color factors are made of contractions of the invariant pairing  and structure constants of the Lie algebra of the gauge group. The numerators $n_i$ include all remaining dependence on polarization vectors $\epsilon_i$ and momenta $p_i$. The integral $\int_l$ is over all internal loop momenta. This decomposition is not unique, and there exist many ways of writing the same amplitude using such cubic graphs. The color-kinematics duality holds if there is some choice of the numerators $n_i$ such that they obey the same identities as the color factors $c_i$. In this paper the relations obeyed by the color factors will be due to the Jacobi identity only, and thus the color-kinematics duality implies
\begin{equation}
    c_i+c_j+c_k = 0 \leftrightarrow n_i +n_j+n_k = 0 \ .
\end{equation}
Having obtained numerators that obey the duality, the on-shell double-copy prescription replaces color factors with an extra copy of the kinematic numerators, $c_i\to \tilde{n}_i$ in the amplitudes, resulting in an amplitude from some gravity-like theory,
\begin{equation}
    \mathcal{M}_{\textrm{double copy}} = \sum_{i\in \Gamma_3}\int_l \frac{n_i \tilde{n}_i}{D_i} \ .
\end{equation}
If the input numerators came from amplitudes which are invariant under the transformation of the external polarizations $\epsilon_\mu\to p_\mu$ then the double-copied amplitude is invariant under linearized diffeomorphisms $\epsilon_\mu \tilde{\epsilon}_\nu \to p_\mu \tilde{\epsilon}_\nu +\epsilon_\mu p_\nu $.
Note that the new numerators may come from a completely different gauge theory.

To see that Chern-Simons theory obeys the duality, we take as a starting point the AKSZ construction of Chern-Simons theory \cite{Alexandrov:1995kv} (see also \cite{Axelrod:1991vq}),
\begin{equation}
    S_{CS} = \int d^3x~ d^3\theta ~\Big(\frac{1}{2}\mathcal{A}^ad\mathcal{A}^b\delta_{ab} + \frac{1}{6}\mathcal{A}^a\mathcal{A}^b\mathcal{A}^c f_{abc}\Big) \ ,
\end{equation}
where $\mathcal{A}^a(x,\theta)$ is interpreted as a Lie algebra-valued superfield in the odd variable $\theta^\mu\equiv dx^\mu$. For further details on the BV construction of Chern-Simons theory see appendix \ref{s:CS}. With this $\theta$ variable we write the exterior derivative as $d=\theta^\mu\partial_\mu$. The indices $a,b,c$ are Lie algebra indices 
 with $\delta_{ab}$ and $f_{abc}$ the invariant pairing and structure constants respectively. 
The action obeys the classical master equation with respect to the BV-symplectic form
\begin{equation}
    \omega_{CS} = \int d^3x~d^3\theta~ \delta\mathcal{A}^a\wedge \delta \mathcal{A}^b\delta_{ab} \ .
\end{equation}

Color-kinematics duality emerges after making the gauge choice $\mathcal{A}^a\in \textrm{Im}(d^\dagger)$, where $d^\dagger=\frac{\partial}{\partial x^\mu}\frac{\partial}{\partial\theta_\mu}$. On this subspace we can define a non-degenerate pairing \cite{Ben-Shahar:2024dju} 
\begin{equation}\label{pairing_CS}
    \langle \omega_1 , \omega_2 \rangle =\int \omega_1 \wedge \xi_2 \ ,
\end{equation}
with $\omega_i=d^\dagger \xi_i$ (note that the pairing can be extended to $\rm{ker}(d^\dagger)$ also). It satisfies
\begin{equation}\label{symmetry_pairing}
\langle \omega_1 , \omega_2 \rangle=(-1)^{(\deg\omega_1+1)(\deg\omega_2+1)}\langle \omega_2 , \omega_1 \rangle\;.
\end{equation}
With this pairing the gauge-fixed Chern-Simons action can be written as
\begin{equation}
    S_{CS} = \frac{1}{2}\langle \mathcal{A}^a, \Box \mathcal{A}^b\rangle \delta_{ab} + \frac{1}{6}\langle \mathcal{A}^a,\{\mathcal{A}^b,\mathcal{A}^c\}\rangle f_{abc}~, 
\end{equation}
where $\Box=dd^\dagger + d^\dagger d$. The bracket appearing in the interaction term is the one generated by $d^\dagger$ on $\Omega(\Sigma_3)$ and reads
\begin{eqnarray}\label{bracket_forms}
\{\omega_1,\omega_2\} =(-1)^{\deg\omega_1}\left( d^\dagger (\omega_1\omega_2) - d^\dagger(\omega_1)\omega_2 - (-1)^{\textrm{deg }\omega_1}\omega_1 d^\dagger \omega_2\right) \;.
\end{eqnarray}
It satisfies the Jacobi identity because $d^\dagger$ satisfies $(d^\dagger)^2=0$  and is a second-order operator, meaning it obeys
\begin{equation}\label{second_order}
d^\dagger(\omega_1\omega_2\omega_3) =     d^\dagger(\omega_1\omega_2)\omega_3 - d^{\dagger}(\omega_1)\omega_2\omega_3 + \textrm{cyclic}(1,2,3) ~,
\end{equation}
so that it defines the graded Lie superalgebra $(\Omega(\Sigma_3)[1],\{,\})$ where with $[1]$ we mean that the degree is shifted, {\it i.e.} a $k$-form $\omega$ has degree $k-1$. We call the Lie subalgebra 
\begin{equation}\label{CS_kin_gLa}
\kin\equiv{\rm Im} (d^\dagger)[1]\subset\Omega(\Sigma_3)[1]\; 
\end{equation}
the kinematic graded Lie superalgebra or kinematic Lie algebra for short.
The bracket is then just the restriction of (\ref{bracket_forms})
\begin{equation}
\label{bracket_im_codiff} 
\{\omega_1,\omega_2\} = (-1)^{\deg \omega_1}d^\dagger(\omega_1\omega_2)\;.
\end{equation}
The pairing \eqref{pairing_CS} is defined on the kinematic Lie algebra: it is symmetric with respect to the shifted degree thanks to (\ref{symmetry_pairing}) and is invariant, {\it i.e.} 
\begin{equation}
   \langle \omega_1 , \{\omega_2,\omega_3\}\rangle  = \langle \{\omega_1,\omega_2\},\omega_3\rangle \ .
\end{equation}

The (shifted) degree $0$ component $\kin_0$ is an ordinary Lie algebra that is isomorphic, using the metric, to the Lie algebra of exact divergenceless vector fields ${\rm Vect}_{div}^0(\Sigma_3)$. We call
\begin{equation}\label{CS_kin}
    \kin_0={\rm Vect}_{div}^0(\Sigma_3)
\end{equation}
the physical part of the Chern-Simons kinematic algebra, and with this notation we will not distinguish between vector fields and one forms.
This Lie algebra and its bracket play an analogous role to the color structure constant $f_{abc}$, and implies that the Feynman rules computed form this action obey the color-kinematics duality. To illustrate the duality we consider the computation of a simple diagram in momentum space. We take external states $a_i(p_i,\theta)$ which are subject to our gauge condition $a_i\in \textrm{Im}(d^\dagger)$, implying $a_i = ip_i^\mu\frac{d}{d\theta^\mu}{f}(\theta)$. We consider this diagram 
\begin{align}
    \begin{tikzpicture}[baseline=-0.2cm,scale=0.6]
\draw[thick] (-1,-1) -- (0,0) -- (-1,1);
\draw[thick] (0,0) -- (1.5,0)--(2.5,1);
\draw[thick] (1.5,0)--(2.5,-1);
\draw (-1.4,-1.3) node {$1$};
\draw (-1.4,1.3) node {$2$};
\draw (2.8,1.3) node {$3$};
\draw (2.8,-1.3) node {$4$};
\end{tikzpicture}
    &= \frac{1}{(p_1+p_2)^2}f^{a_1 a_2 b}f^{b a_3 a_4} \big\langle \big\{\{a_1 ,a_2\},a_3\big\},a_4\big\rangle \\
    &= \frac{1}{(p_1+p_2)^2}f^{a_1 a_2 b}f^{b a_3 a_4} \int  d^3\theta\ \{a_1 ,a_2\}a_3a_4 \ .
\end{align}
In momentum space the integration over $d^3 x$ is replaced by demanding total momentum conservation.
The color factor vanishes under the cyclic sum of any three labels due to the Jacobi identity, and indeed this holds for the kinematic part as well due to the Jacobi identity of the kinematic Lie bracket $\{\ , \ \}$ and the compatibility of the inner product with the bracket.
This observation extends to all other diagrams computed in this gauge and implies color-kinematic duality. 

The Feynman diagrams of the double copy of Chern-Simons theory with itself contain two integrals over odd variables $\theta,{\theta'}$, but both copies of numerators should be taken to have the same momentum dependence. So, for example,
\begin{align}
    \begin{tikzpicture}[baseline=-0.2cm,scale=0.6]
\draw[thick] (-1,-1) -- (0,0) -- (-1,1);
\draw[thick] (0,0) -- (1.5,0)--(2.5,1);
\draw[thick] (1.5,0)--(2.5,-1);
\draw (-1.4,-1.3) node {$1$};
\draw (-1.4,1.3) node {$2$};
\draw (2.8,1.3) node {$3$};
\draw (2.8,-1.3) node {$4$};
\end{tikzpicture} 
    &= \frac{1}{(p_1+p_2)^2} \int d^3\theta ~d^3{\theta'} \ \big(\{a_1 ,a_2\}a_3a_4\big)\big( \{a'_1 ,a'_2\}a'_3a'_4 \big)\ ,
\end{align}
with $a'$ depending on $p$ and ${\theta'}$.
We want to reverse engineer an action whose Feynman rules generate the double-copied amplitudes. Such a proposal was made in \cite{Ben-Shahar:2021zww},
\begin{equation}\label{cs_gaugefixed}
    S = \int d^3x~ d^3\theta~ d^3{\theta'}~ \Big( 
    \frac{1}{2}\A \frac{\D D}{\Box} \A + \frac{1}{6}\A^3\Big)
\end{equation}
where now $\A = \A(x,\theta,{\theta'})$ is a superfield in two odd coordinates $\theta$ and ${\theta'}$, the two differentials are $D=\theta^\mu\partial_\mu$ and $\mathcal{D}={\theta'}^\mu\partial_\mu$, and it is assumed that the field $\A$ is in the kernel of both $D^\dagger = \frac{\partial}{\partial\theta^\mu}\partial_\mu$ and $\gen = \frac{\partial}{\partial{\theta'}^\mu}\partial_\mu$.
As it stands, this action is invariant under two transformations,
\begin{align}\label{cs_BRST}
    Q \A = D \A + \frac{1}{2}\gen \A^2\ , \ \ \ \ \ \ \mathcal{Q} \A = \D \A - \frac{1}{2} D^\dagger \A^2 \ ,
\end{align}
which obey $Q^2=0$, $\mathcal{Q}^2=0$, and $Q\mathcal{Q} + \mathcal{Q}Q = 0$ as consequence 
of second order properties of $D^\dagger$ and $\gen$. These have the interpretation of BRST symmetries if they are dressed with projections onto the kernels of $D^\dagger$ and $\gen$, see Appendix \ref{a:BRST}.

Our task in the next section is to give the BV description of (\ref{cs_gaugefixed}) and find its gravity interpretation. The lessons we learn from Chern-Simons theory will then be applied to other examples, where in some cases the resultant actions will be local.

\section{A BV action for the double copy of Chern-Simons theory}\label{s:BV-double-CS}

The kinematic Lie algebra ${\cal K}$ is endowed with the invariant pairing (\ref{pairing_CS}) and so it is tempting to consider Chern-Simons (CS) theory for the quadratic Lie algebra ${\cal K}$. This intuition points in the correct direction but it does not give the right construction, because it doubles the coordinates on $\Sigma_3$. A related approach, in which the dimension of spacetime is   doubled, has been considered in ref. \cite{Bonezzi:2024dlv} and we will discuss versions of this in Section \ref{s:KS}.  Here we want to find a BV action that after gauge fixing recovers the double-copy action (\ref{cs_gaugefixed}), where space time is intrinsically three dimensional. First, however, we will study CS for $\so(3,\Re)$, which will be an instructive example to understand the key features of the double-copy construction that we consider later on.

\subsection{CS for $\so(3,\Re)$}\label{ss:so(3)}

Many of the properties of the CS double copy can be recognized in standard CS theory when the Lie algebra is $\g=\so(3,\Re)$, see appendix \ref{s:CS} for the general case.
In this particular setting
\begin{equation}
\g = {\rm Im} (\gen)\subset\wedge \g~,
\end{equation}
where $\gen:\wedge\g^\bullet\rightarrow\wedge^{\bullet-1}\g$ is the generator of the Lie bracket,
defined by the property of being second order and $\gen(X)=0$, $\gen(X\wedge Y)=[X,Y]$ for each $X,Y\in\g$. The property that
${\gen}^2=0$ is equivalent to the Jacobi identity. By introducing a basis ${\theta'_a}$ of $\g$, it reads
\begin{equation}\label{3D-oper-SO(3)}
\gen = \frac{1}{2} f_{ab}^c{\theta'_c}\frac{\partial}{\partial {\theta'_a}}\frac{\partial}{\partial {\theta'_b}}\;.
\end{equation}
Actually every finite dimensional Lie algebra $\g$ extends to a Gerstenhaber algebra $\Gerst=\wedge\g$ with generator $\gen:\wedge^{\bullet} \g\rightarrow\wedge^{\bullet-1}\g$, but $\so(3,\Re)$ is special since the analogue of (\ref{pairing_CS}) defines a  nondegenerate invariant symmetric pairing so that, up to an inessential multiplicative constant, 
\begin{equation}\label{so3_pairing_as_integral}
\Tr(ab)= \int  d^3{\theta'}\  a\xi ~,
\end{equation}
where $a,b\in\so(3,\Re)$ and $b=\gen \xi$. 

Although it is a trivial rewriting, it is useful to describe the CS dgLa structure introduced in the previous section in terms of these data.  
The graded commutative algebra $\tilde\dgLa\equiv\Omega(\Sigma_3)\otimes\wedge \g$ is actually bigraded by
\begin{equation}
\dl(\theta)=\dr({\theta'})=1\,,\;\;\;
\dl({\theta'})=\dr(\theta)=0
\end{equation}
where $\theta$ denotes the local odd generators of $\Omega(\Sigma_3)$. We denote the total degree with $d_{L+R}=d_L+d_R$. The generator (\ref{3D-oper-SO(3)}) is extended to $\gen:{\tilde\dgLa}\rightarrow \tilde\dgLa$ as trivial on $\Omega(\Sigma_3)$. Since it is a second order operator it defines a Gerstenhaber bracket on $\tilde\dgLa$. We see that ${\rm Im}(\gen)=\Omega(\Sigma_3)\otimes \so(3,\Re) \subset\tilde\dgLa$. We now compute for each $a,b\in\so(3,\Re)$ and $\omega,\nu\in\Omega(\Sigma_3)$
\begin{eqnarray*}
[\omega\otimes a,\nu\otimes b] &=& (-1)^{|\omega|+1} \gen(\omega\otimes a \cdot \nu\otimes b)  \cr
& &= (-1)^{|\omega|+1+|\nu|}\gen(\omega\nu\otimes ab)=\omega\nu\otimes[a,b]\;,
\end{eqnarray*}
where $d_{L+R}(\omega\otimes a)=|\omega|+1$  and $|\omega|$ denotes the form degree. Analogously, we check the pairing
$$
\langle\omega\otimes a,\nu\otimes b\rangle = \int \omega\otimes a \cdot \nu\otimes \xi (-1)^{|\nu|} = \int \omega\nu\otimes a\xi = \int \omega\nu\ \Tr(ab)\,,
$$
where $\gen\xi = b$. Remark that the de Rham differential $d$ is a derivation of $[,]$ since $[d,\gen]=0$. We conclude that $({\rm Im}D^\dagger,d,[,])$ with shifted degree $\deg = d_L+d_R-1$ coincides with the CS dgLa described in the previous section together with the $\deg=-3$ pairing $\langle,\rangle$.

The superfield $\A$ is now expanded as 
\begin{equation}
\A =\A_0^a {\theta'_a} + \A_{\mu}^a{\theta'_a}\theta^\mu +\frac{1}{2} \A_{\mu\nu}^a{\theta'_a}\theta^\mu\theta^\nu + \frac{1}{6}\A_{\mu\nu\rho}^a{\theta'_a}\theta^\mu\theta^\nu\theta^\rho\;.
\end{equation}
We easily recognize that the field components have the correct ghost degree assignement as in the standard CS superfield if we declare
\begin{equation}
\gh = 1-\deg = 2 -\dl-\dr\;,
\end{equation}
{\it i.e.} by saying that $\A$ has total degree ${\rm tot} =\gh + d_L + d_R= 2$.
We remark that since ${\rm tot}$ is the relevant degree that matters for the definition of the bracket by means of the generator $\gen$ we have that $[\A,\A]=\gen(\A^2)$.

The BV symplectic form is,
\begin{equation}
 \label{so3_CS_symplectic}
 \omega_{BV} =
 \langle \delta\A,\delta \A\rangle\,,
\end{equation}
and the full action can then be written as
\begin{equation}
 S = \frac{1}{2}\langle \A,d\A \rangle + \frac{1}{6} \langle\A,[\A,\A]\rangle=\frac{1}{2}\langle \A,d\A \rangle + \frac{1}{6} \langle\A,\gen \A^2\rangle\;.
\end{equation}
Finally, we note that by using (\ref{so3_pairing_as_integral}) the solution of CME can be written as
\begin{equation}
 \label{so_CS_action}
 S = \int d^3x~d^3\theta ~d^3{\theta'}\ \Big(-\frac{1}{2}\A d\tilde{\A} + \frac{1}{6} \A^3\Big) \;.
\end{equation}
with $\A = \gen \tilde{\A}$ once again.

Here the crucial observation is that the dgLa pairing is expressed as an integral over $T[1]\Sigma_3\times\Re^3[1]$, and thus the measure has total degree $-6$. We show a similar example in Appendix \ref{ss:JT} for JT gravity.

\subsection{BV double copy of Chern Simons in flat space}
\label{dc_CS_flat}
In order to construct the double copy of CS, we will simply modify the construction discussed above by choosing the generator of the kinematic Lie algebra instead of the generator $\gen$ of $\so(3,\Re)$.

We consider here $\Sigma_3=\Re^3$ so that the commutative graded algebra $\Omega(\Sigma_3,\wedge T\Sigma_3)$ can be identified with
\begin{equation}
\tilde{\dgLa} = C^{\infty}(\Re^3)\otimes\wedge \Re^3\otimes\wedge\Re^3
\end{equation}
generated by $C^\infty(\Re^3)$ and $\theta^\mu$, ${\theta'_\mu}$, $\mu=1,2,3$. It is bigraded by
\begin{equation}\label{bigrading_double_copy}
\dl(\theta)=\dr({\theta'})=1\,,\;\;\;
\dl({\theta'})=\dr(\theta)=0\;.
\end{equation}
Let us define the shifted degree
\begin{equation}
    \deg = \dl+\dr-1 \;.
\end{equation}
We extend the de Rham differential to $\tilde{\dgLa}$ as
\begin{equation}
D = \theta^\mu\frac{\partial}{\partial x^\mu}
\end{equation}
and the codifferential as
\begin{equation}
\gen = \frac{\partial^2}{\partial{\theta'_\mu} \partial x^\mu}\;,
\end{equation}
 with the relations\footnote{In what follows we use the notation $[~,~]$ for graded commutator. Thus for odd elements it is defined as anti-commutator.}
\begin{equation}\label{double_copy_algebra}
[D,D]=[\gen,\gen]=[D,\gen] =0~.
\end{equation}
Then the data $(\tilde{\dgLa},D,\gen)$  can be understood as a dgLa with grading given by $\deg$ and Lie bracket defined by the second order operator $\gen$. 
So we can repeat the previous formulation of $\so(3,{\mathbb R})$-CS with a different $\gen$. We consider the dgLA subalgebra
\begin{equation}
\dgLa\equiv{\rm Im} (\gen) 
\end{equation}
with the  pairing defined as
\begin{equation}
 \label{pairing_double_copy}
 \langle a,b\rangle = \int d^3x\ d^3\theta\ d^3{\theta'} \ a(x,\theta,{\theta'}) \xi(x,\theta,{\theta'})~,
\end{equation}
where $b = \gen \xi$.
We get the following decomposition of $\dgLa$ 
in the degree $\dr$
\begin{equation}
\label{dec_im_codif}
\dgLa = \dgLa_0 + \dgLa_1 + \dgLa_2\;,
\end{equation}
so that $\A\in\dgLa$ decomposes as
\begin{equation}
\A (x, \theta, {\theta'}) = {\mathbf B}(x, \theta) + \theta_\mu' {\mathbf A}^\mu (x,\theta)  
+ \theta_\mu' \theta_\nu'{\mathbf C}^{\mu\nu} (x, \theta) ~. 
\end{equation}
We use lower indices for $\theta'$ and upper for $\theta$ coordinates, such that the field components are
\begin{equation}
B_{\mu_1\ldots\mu_r}, A_{\mu_1\ldots\mu_r}^{\nu}, C_{\mu_1\ldots\mu_r}^{\nu_1\nu_2}
\end{equation}
satisfying
\begin{equation}
\partial_\rho A_{\mu_1\ldots\mu_r}^{\rho}=\partial_\rho C^{\rho\nu}_{\mu_1\ldots\mu_r}=0 \ .
\end{equation}
The ghost degree of every component is $2-\dl-\dr$ so that
\begin{equation}
{\rm gh} (B_{\mu_1,\ldots \mu_r}) =2-r,\;\;
{\rm gh} (A_{\mu_1,\ldots \mu_r}^\nu)= 1-r\,,\;\; \gh ( C_{\mu_1,\ldots \mu_r}^{\nu_1\nu_2})=-r\;.
\end{equation}
The general result described at the end of appendix \ref{s:CS}
implies that
\begin{equation}\label{double_copy_BV_action_1}
S_{BV} = \frac{1}{2}\langle \A,D\A\rangle + \frac{1}{6}\langle \A,\gen\A^2\rangle
\end{equation}
solves the CME with respect to BV symplectic structure $\omega_{BV} =\langle \delta \A, \delta \A \rangle$. 

We can now also introduce the de Rham operator in the ${\theta'}$ direction
\begin{equation}
\D = {\theta'}^\mu\frac{\partial}{\partial x^\mu}
\end{equation}
satifsying
\begin{equation}
\square = g^{\mu\nu}\frac{\partial^2}{\partial x^\mu\partial x^\nu}=[\D,\gen] =\D\gen + \gen \D 
\end{equation}
and 
\begin{equation}
[\D,\D]=[D,\gen]=0~.
\end{equation}
  The invariant pairing (\ref{pairing_double_copy}) can be expressed as
\begin{equation}\label{double_copy_pairing_bis}
\langle a,b\rangle = \langle a, \frac{1}{\square} \gen\D b  \rangle =\langle a,  \gen \frac{1}{\square} \D  b  \rangle =\int d^3x\ d^3\theta\ d^3{\theta'}~ a \frac{1}{\square} \D b\;,
\end{equation}
and the action (\ref{double_copy_BV_action_1}) reads
\begin{eqnarray}
 \label{double_copy_BV_action}
S_{BV} = 
\int d^3x\ d^3\theta\ d^3{\theta'} \Big ( \frac{1}{2}\A\frac{\D D}{\square}\A + \frac{1}{6} \A^3 \Big ) \;.
\end{eqnarray}

In deriving (\ref{double_copy_pairing_bis})  we implicitly assumed that $\ker\square\cap{\rm Im}(\D)=0$, {\it i.e.} zero modes vanish. If we consider euclidean signature, this condition is satisfied if we restrict to field configurations vanishing suitably at infinity. In general, one should keep the zero modes as background and expand fields around them. We leave this more careful analysis to future work.

By looking at the non-local form of (\ref{double_copy_BV_action}) it may not be completely obvious that the highly non-local formalism still produces local BV transformations. Indeed, let us check it explicitly. Let $Q_{BV}$ denote the Hamiltonian vector field of $S_{BV}$, {\it i.e.} it satisfies
$\iota_{Q_{BV}}\omega = \delta S_{BV}$. We then compute
$$
\iota_{Q_{BV}}\frac{1}{2}\langle \delta\A,\delta\A\rangle =
\langle Q_{BV}(\A),\delta\A\rangle =\int \frac{1}{\square}\D Q_{BV}(\A)\delta\A = \int \delta\A(\frac{1}{\square}\D D\A + \frac{1}{2} \A^2)
$$
so that
\begin{equation}
\frac{1}{\square}\D Q_{BV}(\A) = \frac{1}{\square}\D D\A + \frac{1}{2} \A^2\;.
\end{equation}
By multiplying both sides by $\gen$ we finally get the local BV transformations
\begin{equation}
\label{double_copy_BV_transformations}
Q_{BV}(\A) = D\A + \frac{1}{2} \gen\A^2\;.
\end{equation}
  Let us now introduce the second-order operator
\begin{equation}
D^\dagger = \frac{\partial^2}{\partial{\theta_\mu} \partial x^\mu}\;,
\end{equation}
so that we can impose the gauge-fixing condition $\A\in \rm{Im}(D^\dagger)$ such that the field is in $\rm{Im}(D^\dagger)\cap\rm{Im}\gen$. It is clear that (\ref{double_copy_BV_action}) evaluated on configurations that satisfy the gauge-fixing condition reproduces the double-copy action (\ref{cs_gaugefixed}). 

The BV operator $Q_{BV}$ does not preserve the gauge-fixing condition so it does not automatically define a symmetry of $\rm{Im}(D^\dagger)\cap\rm{Im}\gen$, however, one can dress it with a projection onto the gauge-fixed subspace and the resulting vector field is a symmetry of the gauge-fixed action that squares to zero on shell \cite{Bonechi_2016}. This recovers the first of the two BRST transformations of (\ref{cs_BRST}). In order to understand the second one, let us point out that the gauge-fixed action is now symmetric in the exchange of the two sides of the double copy. The constraint $\A\in \rm{Im}(\gen)$ can then be viewed as a gauge-fixing condition of a different BV theory, whose BV operator fixes the second BRST transformation in \eqref{cs_BRST}. We will discuss this in more detail in Appendix \ref{a:BRST}.

Let us finally describe the structure of the underlying classical gauge theory. The physical fields are the fields of ghost degree $0$, {\it i.e.}
\begin{equation}\label{classical_fields_dCS}
B\equiv B_2\in\Omega^2(\Re^3),\ \ \ \
A\equiv A_1\in\Omega^1(\Re^3,T\Re^3),\ \ \ \
C\equiv C_0\in\Gamma(\wedge^2 T\Re^3)
\end{equation}
and by restricting (\ref{double_copy_BV_action}) to them we get the classical action
\begin{align}\label{classical_action_dCS}
S_{c\ell} =& \int B\frac{1}{\square}\D D C + \frac{1}{2}A\frac{1}{\square}\D D A + \frac{1}{6}A^3 + A B C\cr
=&
\int d^3x \ \epsilon^{\mu\nu\rho}\epsilon_{\lambda\tau\sigma}(\frac{1}{2}A_\mu^\lambda\frac{1}{\square}\partial_\nu \partial^\tau A_\rho^\sigma -B_{\mu\nu}\frac{1}{\square} \partial_\rho\partial^\lambda C^{\tau\sigma} \cr
&  \;\;\;\;\;\;\;\;\;\;\;\;- \frac{1}{6}A_\mu^\lambda A_\nu^\tau A_\rho^\sigma+ A_\mu^\lambda B_{\nu\rho}C^{\tau\sigma})\;.
\end{align}
 This action implies the following local equation of motions 
\bea\label{def-eqs}
 &&  \partial_{\mu} B_{\nu\rho} + A^\lambda_{\mu} \partial_\lambda  B_{\nu \rho} + {\rm cyclic}(\mu, \nu, \rho) =0~,\label{eqom_MC_A-0}\\
 &&  \partial_{[\mu} A_{\nu]}^\rho + A^\lambda_{[\mu} \partial_\lambda A^\rho_{\nu]} + 4 C^{\rho\sigma} \partial_\sigma B_{\mu\nu}=0~,\label{eqom_MC_A}\\
 && \partial_{\rho} C^{\mu\nu}  + A^\sigma_{\rho} \partial_\sigma C^{\mu\nu} + C^{\nu\sigma} \partial_\sigma A^\mu_{\rho} - C^{\mu\sigma} \partial_\sigma A^\nu_{\rho} =0~~,\label{eqom_MC_A-1}
  \eea 
where anti-symmetrization of indices does not include a factor of $1/2$. The ghosts are
$B_1\in\Omega^1(\Re^3)$ and $A_0\in\Omega^0(\Re^3,T\Re^3)$.
The gauge transformations are read from the terms of (\ref{double_copy_BV_transformations}) that are linear in the ghosts. They are parametrized by $\beta=\beta_\mu\theta^\mu,\alpha=\alpha^\mu{\theta'_\mu}$, satisfying $\partial_\rho\alpha^\rho=0$ and read
\begin{eqnarray}
\label{double_copy_gauge_transf_1}
\delta B &=& D\beta + \gen (\beta A + \alpha B) \cr
\delta A &=& D\alpha +\gen(\beta C + \alpha A)\cr
\delta C &=& \gen(\alpha C)
\end{eqnarray}
or in components  
\begin{eqnarray}
\label{double_copy_gauge_transf_2}
\delta A_\mu^\nu &=& \partial_\mu\alpha^\nu + \partial_\lambda(\alpha^\lambda A^\nu_\mu - A^\lambda_\mu \alpha^\nu) +\partial_\lambda(\beta_\mu C^{\lambda\nu})~, \cr
\delta B_{\mu\nu}&=& (d\beta)_{\mu\nu} +  ( A_\mu^\lambda \partial_\lambda\beta_\nu -A_\nu^\lambda\partial_\lambda\beta_\mu)+\alpha^\lambda\partial_\lambda B_{\mu\nu}~, \cr
\delta C^{\mu\nu} &=& \partial_\lambda( \alpha^\lambda C^{\nu\mu}+\alpha^\mu C^{\lambda\nu}+\alpha^\nu C^{\mu\lambda})~.
\end{eqnarray}
The transformations \eqref{double_copy_BV_transformations} with gauge parameter $\gamma = \beta+\alpha$ obey the algebra
\begin{equation}
    [\delta_{\gamma_1},\delta_{\gamma_2}] = \delta_{[\gamma_1,\gamma_2]} \ ,
    \hspace{2cm} [\gamma_1,\gamma_2] = \frac{1}{2}\gen(\gamma_1\gamma_2) \ ,
\end{equation}
which in components gives
\begin{equation}
[\alpha_1,\alpha_2] = \frac{1}{2}\gen(\alpha_1\alpha_2)\,,\;
[\alpha,\beta] = \frac{1}{2}\gen(\alpha\beta)=\frac{1}{2}\iota_\alpha(d\beta)~,~~
[\beta_1,\beta_2]=0\;.
\end{equation}
This is the semidirect product
\begin{equation}
\kin_0\ltimes \Omega^1(\Re^3)
\end{equation}
of the Chern Simons kinematic Lie algebra $\kin_0$ (\ref{CS_kin}) of exact divergenceless vector 
fields with its representation on one forms given above.

\subsection{A gravity interpretation of CS double copy}

Let us concentrate on the physical fields and for the moment let us set $B=0$ and $C=0$. 
The equations of motion (\ref{eqom_MC_A-0})-(\ref{eqom_MC_A-1}) then become 
\be\label{MC-only-A}
\partial_{[\mu} A_{\nu]}^\rho + A^\lambda_{[\mu} \partial_\lambda A^\rho_{\nu]} =0
\ee   
together with the condition $\partial_\mu A^\mu_\nu=0$. Our claim is that the above Maurer–Cartan equation parametrizes deformations of a flat metric $g = \eta + \delta g $ around the canonical flat metric $\eta$ with fixed volume form $\det g =1$
  \begin{equation}
      R_{\mu\nu}(\eta+ \delta g)=0~,~~~\det (\eta +\delta g) =1~. 
  \end{equation}
We will first show that our equations of motion match those obtained in a certain sector of 3d Gravity, and then explain how to directly extract the geometric quantities form our fields.

To show that our equations of motion solve a certain sector of 3d equations of motion, recall that Ricci flatness in 3d  is equivalent to Riemann flatness. Let us use the first-order formalism with
the vielbein 1-form $e^a_\mu$ a and a spin connection $\omega_{\mu b}^a$ where metric is written as follows
\be
  g_{\mu\nu}= e_\mu^a \eta_{ab} e^b_\nu ~,
\ee
here we use the Latin letters for flat indices and Greek letters for the curved indices. Let us introduce the dual vector frames
\begin{equation}
    E^\mu_a e^a_\nu = \delta^\mu_\nu~,~~~~~~
    E^\mu_a e^b_\mu = \delta^b_a
\end{equation}
so that the inverse metric can be written as follows
\begin{equation}
    g^{\mu\nu} = E^\mu_a \eta^{ab} E^\nu_b~.
\end{equation}
In the first-order formalism the condition $R_{\mu\nu}(g) =0$ is equivalent to the following two equations 
\begin{eqnarray}
    de^a+ \omega^a_b \wedge e^b=0~,\\
    d\omega^a_b + \omega^a_c \wedge \omega^c_b=0 ~,
\end{eqnarray}
where we always assume that $\det e \neq 0$. 
Let us now define the connection $\nabla_\mu$ which annihilates $e$ (or $E$)
\begin{equation}
    \nabla_\mu e^a_\nu = \partial_\mu e^a_\nu + \omega_{\mu b}^a e^b_\nu -
    \Gamma^\rho_{\mu\nu} e_\rho^a=0~.
\end{equation}
This defines $\Gamma$ in terms of the spin connection $\omega$
\begin{equation}
    \Gamma^\rho_{\mu\nu} = E^\rho_a \Big ( \partial_\mu e^a_\nu + \omega_{\mu b}^a e^b_\nu \Big )~. 
\end{equation}
Due to one of the equations $\Gamma$ is the Levi-Civita connection 
\begin{equation}
    \Gamma^\rho_{\mu\nu}= \Gamma^\rho_{\nu\mu}~~\longleftrightarrow~~de^a+ 
    \omega^a_b \wedge e^b=0~.
\end{equation}
On $\mathbb{R}^3$ we can solve the second equation by setting $\omega=0$. 
Thus 
two equations collapse to one simple equation
\begin{equation}
    d e^a=0~,
\end{equation}
with the additional condition that $\det e \neq 0$. This simple equation for $e$ implies the following two equations for $E$
\begin{eqnarray}
\{ E_a, E_b \}=0~,\label{Lie-bracket-frame}\\
\partial_\mu \Big ( (\det e) E^\mu_a \Big )=0~,\label{cov-diver}
\end{eqnarray}
where $\{~,~\}$ is Lie bracket of vector fields. 
The equation (\ref{Lie-bracket-frame}) follows from the relation 
\begin{equation}
\{ E_a, E_b\}^\nu e_\lambda^a e_\gamma^b = E^\nu_a (\partial_\gamma 
e^a_\lambda - \partial_\lambda e^a_\gamma)~.
\end{equation}
The second equation (\ref{cov-diver}) follows from algebraic relations 
\begin{eqnarray}
    \epsilon_{\mu\nu\rho} \det e = \epsilon_{abc}~ e^a_\mu e^b_\nu e^c_\rho~,\label{1-eq-forE}\\
    (\det e) E^\mu_a = \frac{1}{2}  \epsilon^{\mu\nu\rho} \epsilon_{abc} e^b_\nu e^c_\rho~, \label{2-eq-forE}
\end{eqnarray}
and $de^a=0$. Remember that $\det e = \sqrt{g}$ and thus the equation (\ref{cov-diver}) is saying that the vector frames $E$ are divergenceless 
with respect to the volume form defined by the metric $g$. The equations (\ref{Lie-bracket-frame}) and (\ref{cov-diver}) are equivalent to (\ref{MC-only-A}) upon the identification
\begin{equation}
    E^\mu_a = \delta^\mu_a + A^\mu_a
\end{equation}
and thus we can identify flat and curved indices, and $E$ is invertible when $A$ is assumed to be small. Moreover, we assume that $\det e=1$. Our previous discussion guarantees that the deformed metric remains flat. 
The relation between metric $g$ and $A$ is non-linear. The inverse metric is related to $A$ as follows
\begin{equation}
    g^{\mu\nu} = \eta^{\mu\nu} + A^{\mu\nu} + A^{\nu\mu} + A^\mu_\rho A^{\nu\rho}~,
\end{equation}
where we raise and contract indices using the flat canonical metric $\eta$. In this work our main interest is in $\mathbb{R}^3$, but the present discussion can be generalized to deformations of a flat metric around a given flat metric upon the covariantization of some of the formulas by using a flat connection. 
We can consider a more general smooth manifold $\Sigma_3$ so that $\theta$ and $\theta'$ are only local coordinates and define $D$ and $\gen$ by using a flat affine connection. It is rather straightforward to check that (\ref{double_copy_algebra}) implies that the connection is the Levi Civita connection of a flat metric.  
 The general treatment of flat manifolds would require addressing the problem of zero modes which we leave aside for now.

Next we will show that indeed this interpretation of the double copy fields is natural. As before we treat $A$ as a fluctuation of another field, $E_a^\mu = \delta_a^\mu + A^\mu_a$, which will be interpreted as an inverse vielbein with equations of motion,
\begin{equation}
    E_{[a}^\nu \partial_\nu E_{b]}^\mu = 0 \ .
\end{equation}
Let us now suppose that $E$ satisfies these equations of motion and add to it a small deformation $E\to E + \epsilon$, then the equation of motion for the deformation can be written as
\begin{equation}
    E^\mu_{[a}\big(\partial_\mu \epsilon^\nu_{b]} + (\partial_\rho e^c_\mu)E^\nu_c \epsilon^\rho_{b]}\big)  + \epsilon_{[a}^\nu \partial_\nu \epsilon_{b]}^\mu = 0 \ ,
\end{equation}
with $e^a_\mu$ the inverse of $E_a^\nu$.
The first term is the covariant derivative of the deformation with the connection given by
\begin{equation}
    E_a^\rho \partial_\mu e_\nu^a \equiv \Gamma^\rho_{\mu\nu} \ .
\end{equation}
This connection exactly agrees with the standard Levi-Civita connection for the metric $g_{\mu\nu} = e_\mu^a e^b_\nu \eta_{ab}$ on the support of the equations of motion. Plugging it into the definition of the Riemann curvature tensor directly yields 
\begin{equation}
    R_{\mu\nu\rho\sigma} = 0 \ .
\end{equation}
If we restore the $C$ and $B$ fields it is convenient to introduce the Courant geometry and study its flat deformations. This subject lies outside of the scope of this paper, but in Appendix \ref{a:deformations} we provide a geometrical derivation of the equations (\ref{eqom_MC_A-0})-(\ref{eqom_MC_A-1}) and we leave further geometrical details for future work.

\section{Double copy of formal CS theory}\label{s:formal-CS}

\subsection{The kinematic Lie algebra}
We saw in ref. \cite{Ben-Shahar:2024dju} that a second order $D^\dagger$ implies off-shell color-kinematics duality, we will review this construction in this subsection and introduce notation for subsequent sections.

Consider the commutative differential graded algebra $(\cdga, D)$ with graded commutative multiplication and differential $D: \cdga^\bullet \rightarrow \cdga^{\bullet+1}$, together with non-degenerate integration $\int d\mu: \cdga \rightarrow \Re$ of degree  $-3$ satisfying
\bea
 \int d\mu \ Da = 0~.
\eea
   
We assume that $\cdga$ is bounded, {\it i.e.} it has the following structure,
\begin{equation}\label{formal-CS-degree}
\begin{tikzcd}
 & \cdga^{-r} \arrow[r,"D"] & \cdga^{-r+1}  \arrow[r, "D"]  &    ....   \arrow[r, "D"]       & \cdga^{r+2}     \arrow[r, "D"]   & \cdga^{r+3}
\end{tikzcd}
\end{equation}
for some non negative integer $r$. Let $\g$ be a quadratic Lie algebra with nondegenerate pairing $\Tr$ and let us consider the dgLa $(\cdga\otimes\g,D,[,])$ with $\deg=-3$ pairing defined as $\langle a\otimes X,b\otimes Y\rangle=\int d\mu\ ab \Tr(XY)$. Let us suppose that $\g$ is compact and let us fix an orthonormal basis of $\g$. The formal Chern-Simons theory takes the usual form
\bea\label{BV-general-CS}
    S_{CS} = \int d\mu \left(\frac{1}{2} \A^a D \A^b\delta_{ab}+\frac{1}{6} \A^a \A^b\A^c f_{abc} \right) \ ,
\eea 
see also appendix \ref{s:CS} for the standard definition of Chern-Simons theory.
The superfield now decomposes as 
\begin{equation}
\A=A_{-r}+ A_{-r+1}+\ldots A_{r+3}
\end{equation}
where $\gh(A_i) = 1-i$.
The BV symplectic form is given by 
\bea 
    \omega = \frac{1}{2}\int d\mu \ \delta \A^a\wedge \delta \A^b\delta_{ab} ~.
\eea
Color-kinematics duality holds if there exists a linear operator $D^\dagger:\cdga^\bullet\rightarrow\cdga^{\bullet-1}$ that is second order, satisfies $D^\dagger{}^2=0$ and
\begin{equation}\label{properties_D+}
\int d\mu ~D^\dagger (\omega) \nu =   (-1)^{\deg \omega} \int d\mu~ \omega D^\dagger (\nu)~
\end{equation}
for each $\omega,\nu\in\cdga$.
 Moreover, the degree $0$ operator
\begin{equation}
\square =D D^\dagger + D^\dagger D
\end{equation}
in all the examples that we are going to describe will be the Laplace  operator.  
Then the gauge fixing is defined as $\A\in{\rm Im}D^\dagger$ (where $D^\dagger$ is extended to $\cdga\otimes\g$).

The second-order condition (see equation \eqref{second_order}) for $D^\dagger$ defines the kinematic graded Lie superalgebra $\kin= {\rm Im}(D^\dagger)\subset\cdga$ together with the invariant pairing
\begin{equation}
\langle\omega,\nu\rangle =\int d\mu\ \omega \xi
\end{equation}
where $\nu=D^\dagger\xi$.

\subsection{Double copy of formal CS in flat space}\label{double_copy_formal}

  For simplicity we will assume that $(\cdga,D)$ is the ring of functions of a $Q$-manifold $(\E,D)$ with body a smooth manifold $\Sigma_d$, of dimension $d$. Furthermore we assume that the nonvanishing components of $\cdga$ have odd degree (in short, $\E$ is a NQ manifold). Although this is not the most general setting, it is enough to include all the examples discussed in the following Sections.  

We consider the flat case $\Sigma_d=\Re^d$ so that $\E$ has global coordinates $x^\mu$ of degree zero and $\theta^a$ of odd degree (possibly negative). The $\theta^a$ freely generate a commutative graded algebra that we denote by $\cdga_*$. Furthermore we assume that $D$ and $D^\dagger$ are constant differential operators. We assume that the measure of degree $-3$ factors as $d\mu = d^d x~d^k\theta$. 

Let us consider two such $\E$ and $\E'$, possibly non isomorphic, with the same body $\Sigma_d$. We can now repeat the same construction of Section \ref{dc_CS_flat}, let us consider the commutative graded algebra 
\begin{equation}\label{ordinary_dc_space}
\cdga(\E,\E')=C^\infty(\Re^d)\otimes \cdga_*\otimes \cdga_*'\;.
\end{equation}
We denote with $\dl$ the degree on the first factor and $\dr$ the degree on the second. We denote with the same symbol $D$ and $D^\dagger$  the (trivial on the second factor) extension to $\cdga(\E,\E')$ and with $\D$ and $\gen$ the (trivial on the first factor) extension of $D'$ and $D^\dagger{}'$. We have clearly that $D$ and $\D$ are first order and $D^\dagger$ and $\gen$ are second order. Moreover, since these four operators are constant (meaning independent of $x^\mu$) they satisfy
\begin{equation}
D\gen+\gen D=0\;.
\end{equation}
If we denote with $[\ ,\ ]$ the bracket defined by $\gen$, then $\tilde\dgLa\equiv(\cdga(\E,\E',D,[,])$ endowed with the degree $\deg=\dl+\dr-1$, is a dgLa. We consider the dgLa subalgebra
\begin{equation}
\dgLa = \gen\tilde{\dgLa}
\end{equation}
together with its pairing
\begin{equation}
\langle \omega,\nu\rangle = \int d^d x ~d^k\theta~ d^k{\theta'}\ \omega(x,\theta,{\theta'})\xi(x,\theta,{\theta'})
\end{equation}
where $\nu=\gen\xi$. Now the superfield $\A$ is expanded in the component fields $A_{i,j}$ 
according to the $\dr$ and $\dl$ degrees. The ghost degree is 
\begin{equation}
\gh(A_{i,j}) = 2-i-j\;.
\end{equation}
The action 
\begin{equation}\label{double_copy_BV_action_general}
S_{BV} = \frac{1}{2}\langle \A,D\A\rangle + \frac{1}{6}\langle \A,\gen\A^2\rangle \ ,
\end{equation}
as defined in (\ref{double_copy_BV_action_1}), satisfies the master equation with the analogous symplectic form. The classical fields are now
$A_{2-j,j}$ with $-r\leq j \leq 2+r$.
The ghosts are $A_{1-j,j}$ with $-r\leq j \leq 1+r$. As a consequence the gauge algebra of the physical action $S_{phys}=S_{BV}(A_{2-j,j})$ always contains the physical part of the kinematic algebra as a subalgebra. This action can be further gauge fixed by requiring that the fields are in $\textrm{Im}(D^\dagger)$, in which case the two BRST symmetries \eqref{cs_BRST} discussed for Chern Simons appear. We remind the reader this is because the fully gauge-fixed action treats both sides of the double copy democratically, so one can view either the $D^\dagger$ or the $\gen$ constraint as a gauge-fixing condition.

Finally we point out that in general the double copy action $S_{BV}$ is a non local action, as in the case of CS. Nevertheless we will see in the following sections some examples where this action is actually local.

\section{BF theory in $4d$}\label{s:BF}
In \cite{Ben-Shahar:2024dju} we proved that BF theory in $4d$ satisfies off shell color-kinematic duality. We first review this and extend the results in \cite{Ben-Shahar:2024dju}, to obtain a family of kinematic algebras. Following the notations of section \ref{double_copy_formal}, BF theory in $4d$ corresponds to choosing $\E=T[1]\Sigma_4\times\Re[-1]$ with differential 
\begin{equation}
D=\theta^\mu\frac{\partial}{\partial x^\mu}
\end{equation}
where $\theta^\mu\equiv dx^\mu$ and $\zeta$ are the coordinates of degree $1$ and $-1$ respectively.
The underlying complex ${\cal C}=\Omega(\Sigma_4)[\zeta]=\oplus_{i=-1}^4{\cal C}^i$ is then
\begin{equation}\label{DW_dGA_CSform}
\begin{tikzcd}[sep=small]
 & \Omega^0 (\Sigma_4) \arrow[r] & \Omega^1 (\Sigma_4) \arrow[r] & \Omega^{2}(\Sigma_4) \arrow[r]    & \Omega^3(\Sigma_4) \arrow[r]    & \Omega^4(\Sigma_4)   \\
   \Omega^0  (\Sigma_4) 
   \arrow[r]   &   \Omega^1 (\Sigma_4)  
   \arrow[r]   & \Omega^{2} (\Sigma_4)  \arrow[r] 
   &  \Omega^3 (\Sigma_4)  \arrow[r] 
   &  \Omega^4 (\Sigma_4)   
   &\;.
     \end{tikzcd}
        \end{equation}
If we denote the superfield as $\A=\A(x,\theta,\zeta)$ then the abstract Chern-Simons form of the action of BF theory \eqref{BV-general-CS} is,
\begin{equation}
S_{BF} = \int d^4x~d^4\theta~ d\zeta~ \Big( \frac{1}{2}\A^a D \A^a\delta_{ab} + \frac{1}{6}\A^a\A^b\A^c f_{abc} \Big) \ .
\end{equation}
There is a family of gauge-fixing operators given by,
\begin{equation}
D^\dagger_\lambda= d^\dagger + \lambda \zeta\,,
\end{equation}
which satisfies (\ref{properties_D+}) for all $\lambda\in\Re$
where $d^\dagger$ is the codifferential of an arbitrary metric acting trivially on $\zeta$.
Let us consider first the case $\lambda=0$ that was already discussed in \cite{Ben-Shahar:2024dju}. 
The physical part of the kinematic Lie algebra is isomorphic to
\begin{equation}
\kin_0 = {\rm Vect}_{div}^{ex}(\Sigma_4)\ltimes d^\dagger(\Omega^3(\Sigma_4))
\end{equation}
that is the semidirect product of exact
divergenceless vector fields ${\rm Vect}_{div}^{ex}(\Sigma_4)$ together with their action on coexact two forms by Lie derivative. In the following we will not distinguish the Lie algebra  ${\rm Vect}_{div}^{ex}(\Sigma_4)$ from the isomorphic Lie algebra defined on coexact one forms  $d^\dagger(\Omega^2(\Sigma_4))$ by the codifferential $d^\dagger$.

If $\lambda\not=0$ then the gauge-fixing condition can be solved algebraically. Indeed,
\begin{equation}\label{BF4_homotopy}
\kappa_\lambda = \frac{1}{\lambda} \frac{\partial}{\partial\zeta}
\end{equation}
satisfies
\begin{equation}
[\kappa_\lambda,D^\dagger_\lambda] = 1\;
\end{equation}
so that ${\rm Im}D^\dagger_\lambda = {\ker} D^\dagger_\lambda$ and $\omega+\zeta \nu \in \ker D^\dagger_\lambda$ if and only if 
\begin{equation}\label{image_d+_BF}
\omega +\zeta\nu= D^\dagger_\lambda\kappa_\lambda (\omega+\zeta\nu)= \frac{1}{\lambda}D^\dagger_\lambda \nu = \frac{1}{\lambda}d^\dagger \nu+\zeta \nu~.
\end{equation}
 Since $D^\dagger_\lambda$ is second-order it defines a Lie bracket on ${\rm Im}(D^\dagger_\lambda)$
  which can be written as follows
 \begin{equation}\label{4d-big-bracket}
      {(-1)^{|\nu_1|-1}} D_\lambda^\dagger \Big ( (\frac{1}{\lambda}d^\dagger \nu_1+\zeta \nu_1) (\frac{1}{\lambda}d^\dagger \nu_2+\zeta \nu_2) \Big )  
  = \frac{1}{\lambda^2} d^\dagger ( \{\nu_1, \nu_2 \})  + \frac{1}{\lambda} \zeta 
      \{\nu_1, \nu_2 \} 
 \end{equation}
 where we introduce the notation
\begin{equation}\label{def-BF-bracket}
  \{\nu_1, \nu_2 \} = {(-1)^{|\nu_1|-1}} \big( d^\dagger \nu_1 d^\dagger \nu_2 + (-1)^{|\nu_1|} d^\dagger ( (d^\dagger \nu_1 )\nu_2)
     - d^\dagger (\nu_1 d^\dagger \nu_2)\big)~.  
\end{equation}
 As a consequence of the Jacobi identity for the bracket (\ref{4d-big-bracket}) we can obtain 
 the Jacobi identity for the bracket $\{~,~\}$. The new bracket $\{\nu_1,\nu_2\}$ maps 
 $|\nu_1|$-form and $|\nu_2|$-form to $(|\nu_1|+|\nu_2|-2)$ form. 
 Moreover for this new bracket we can introduce the invariant symmetric pairing  
  \be\label{pairing_local_BF4}
  \langle \nu_1, \nu_2 \rangle\equiv \int \nu_1 \nu_2 \;.
 \ee
 The physical part of the kinematic Lie algebra is now isomorphic to
\begin{equation}
{\cal K}_{\lambda 0} = \Omega^2(\Sigma_4)~,
\end{equation} 
 which are unrestricted two forms, and the bracket and the pairing are closed within two forms.

 If  $\lambda \neq 0$ then the corresponding kinematic algebras are isomorphic and without loss of 
  the generality we can always set $\lambda=1$ by rescaling $\zeta$. In what follows we will set 
   $\lambda=1$. 
  However the relation between 
   the kinematic algebras for $\lambda=0$ and $\lambda \neq 0$ is unclear.

\subsection{Double copy BF $\otimes$ BF}

 Let us construct the double copy theory for 4d BF theory.
We choose $\Sigma_4=\Re^4$ with the euclidean metric. We double the odd coordinates: one set 
 is $(\theta, \zeta)$ with $d_L(\theta)=1$, $d_L(\zeta)=-1$, $d_R(\theta)=0$, $d_R(\zeta)=0$
 and with the operators $D=d=\theta^\mu \partial_\mu$ and $D^\dagger = d^\dagger + \zeta$.  
 Another set of odd coordinates is $(\theta', \zeta')$ with $d_L(\theta')=0$, $d_L(\zeta')=0$,
  $d_R(\theta')=1$, $d_R(\zeta')=-1$ and  with the operators $\D=d'=\theta'^\mu\partial_\mu$
  and $\gen=d'^\dagger+\zeta'$. 
  We can apply the double-copy construction using these operators. If we assume that $\A \in {\rm Im}(\gen)$ then the solution of the CME is
\begin{equation}\label{BV-BF-general}
  S_{BV} =  \int d^4x~d^4\theta~ d\zeta~ d^4{\theta'}~d\zeta' \Big (\frac{1}{2} \A \frac{d d'}{\Box}\A
   + \frac{1}{6} \A^3 \Big )~, 
\end{equation}
 with respect to the following symplectic form
 \begin{equation}
     \omega_{BV} =\int d^4x~d^4\theta~ d\zeta~ d^4{\theta'}~d\zeta'~ \delta \A \frac{d'}{\Box} \delta \A~.
 \end{equation}
 Thanks to the homotopy $\kappa_\lambda$ defined in (\ref{BF4_homotopy})  the double copy construction is now local. Let us show this. 
   Due to (\ref{image_d+_BF}) the superfield $\A \in {\rm Im}(\gen)$ can be expressed as follows
\begin{equation}
\A = \gen \tilde{\A}= {d'}^\dagger \tilde{\A} + {\zeta'} \tilde{\A}~,
\end{equation}
where $\tilde{\A}$ does not depend on ${\zeta'}$.
We can further specify
 \begin{equation}
    \tilde{\A} (x, \theta, \theta', \zeta) = \mathbf{C} (x, \theta, \theta') + \zeta \PHI (x, \theta, \theta')~.
 \end{equation}
  Here the superfield $\mathbf{C}$ is bi-degree $(1,2)$ and  the superfield $\PHI$ is bi-degree $(2,2)$. 
  Now let us evaluate the action (\ref{BV-BF-general}) using this description of ${\rm Im}(\gen)$
 \begin{align}
\label{double_copy_action_BF4d}
S_{BV} &= \frac{1}{2}\int d^4x~d^4\theta~ d\zeta~ d^4{\theta'}~(\tilde{\A} d \tilde{\A} + 
\tilde{\A} ({d'}^\dagger\tilde{\A})^2) \cr
&= \int d^4x~d^4\theta~ d^4{{\theta'}}~ \left[\PHI d\C + \frac{1}{2} \left(\PHI ({d'}^\dagger \C)^2 + 2 {d'}^\dagger(\C{d'}^\dagger\C)\right)\right]\cr
& =\int d^4x~d^4\theta~ d^4{{\theta'}}~ \left[\PHI d\C -  \frac{1}{2} \PHI \{ \C, \C\} \right]~,
\end{align}
 where the Lie bracket $\{~,~\}$  is defined in (\ref{def-BF-bracket}), but here it is defined for the primed coordinates with $d'{}^\dagger$.   The symplectic form can be rewritten as follows
 \begin{equation}
 \label{double_copy_symplectic_BF}
 \omega = 
  \int d^4x~d^4\theta~ d\zeta~ d^4{\theta'} ~ \delta\tilde{\A}\  \delta\tilde{\A}= 2 \int d^4x~d^4\theta~ d^4{{\theta'}}~ \delta\PHI\delta\C\;.
\end{equation}
 We can see that the double copy theory is local. Although it looks like BF theory with respect to an infinite dimensional Lie algebra given by the bracket $\{~,~\}$
   defined in (\ref{def-BF-bracket}), we emphasize that this is not the correct interpretation, because the kinematic algebra acts on the spacetime itself as opposed to an extra copy of the spacetime manifold. We can expand the superfields $\C$ and $\PHI$ in components
 such that the components  $C_{i j}$ and $\Phi_{i j}$ having $\dl$ and $\dr$ degrees equal to $i$ and $j$ respectively (so they are counting the number of $\theta$ and $\theta'$ in the expansion). 
The assignment of the ghost degree is now
$$
\gh(C_{ij}) = 3 -i-j\,,\;\;\; \gh(\Phi_{ij}) = 4-i-j
$$
with $i,j=0,\ldots,4$. The physical fields for $C_{ij}$ correspond to $i+j=3$ and for $\Phi_{ij}$
 with $4=i+j$. It would be interesting to understand the gravitational nature of this higher form theory. 
 
 If we choose $\lambda=0$ then we have a non local double copy action that has some similarities to the Chern-Simons double copy we encountered before. In fact the equations of motion \eqref{eqom_MC_A-0}-\eqref{eqom_MC_A-1} are contained in this double copy, and are supplemented by additional equations for fields that essentially act as Lagrange multipliers for them. The choice of $\gen$ can be viewed as a gauge choice once some $D^\dagger$ is fixed, therefore we expect that the non-local version of the double copy and the local one \eqref{double_copy_action_BF4d} should carry the same physical data. 
It would be interesting to explore if there exists a non-local field redefinition relating the double copies corresponding to different choices of gauge fixing operator.

\section{Yang-Mills theory in 2d}\label{s:2DYM}

In this Section we discuss the double copy construction for 2d YM theory.
 We closely follow the description and notations from our previous work
\cite{Ben-Shahar:2024dju}. Let us start by recalling the BV construction for 2d YM theory. It can be cast in the CS form by choosing $\E = T[1]\Sigma_2\times\Re[1]$ with differential
\begin{equation}
D = d+  {\rm vol} \frac{\partial}{\partial\zeta} = \theta^\mu\frac{\partial}{\partial x^\mu} +  {\rm vol} \frac{\partial}{\partial\zeta} ~, 
\end{equation}
where $x^\mu,\theta^\mu\equiv dx^\mu$ are the coordinates of $T[1]\Sigma_2$ and $\zeta$ is of degree $1$, and $\vol$ denotes a fixed volume form in $\Sigma_2$.
 If we introduce Lie algebra valued superfields $\A^a(x, \theta, \zeta)$
  then the BV action can be written as
  \begin{equation}
     S_{BV} = \int d^2x~d^2\theta~d\zeta~\Big ( \frac{1}{2} \A^a D \A^b \delta_{ab} + \frac{1}{6} \A^a \A^b \A^c f_{abc} \Big )~,  
  \end{equation}
   where $\eta_{ab}$, $f_{abc}$ are Lie algebra data. Let us restrict to the case when $\Sigma_2 = \mathbb{R}^2$ and we choose the canonical volume form
${\rm vol } = \frac{i}{2} \theta^{z} \theta^{\bar{z}} = \frac{i}{2} \theta \bar{\theta}$ assuming the standard 
 complex coordinates on $\mathbb{R}^2= \mathbb{C}$. There exists a gauge fixing so that 2d YM theory obeys 
   off-shell color-kinematic duality. The gauge-fixing operator is given by the following second-order operator 
\begin{equation}\label{codiff_YM2}
D^\dagger = 2\partial^\dagger + \mathcal{I} \frac{\partial}{\partial\zeta} = 
-2 \frac{\partial^2}{\partial \theta \partial\bar{z}} + i \Big ( \frac{\partial }{\partial\bar{\theta}} \bar{\theta} - \bar{\theta} 
\frac{\partial}{\partial \bar{\theta}} \Big ) \frac{\partial}{\partial \zeta} ~,
\end{equation}
  where operations satisfy $\mathcal{I}^2=-1$ and
  $[\partial^\dagger,\mathcal{I}]=0$. Here we have the operator
  \begin{equation}
      \kappa = - {\cal I} \zeta~,
  \end{equation}
  which obeys
   \begin{equation}\label{2DYM-sol-GF}
      [\kappa, D^\dagger]= 1 
   \end{equation}
   and therefore ${\rm Im}(D^\dagger)=\ker(D^\dagger)$ and the gauge-fixing condition can be solved algebraically. 
 Let  ${\cal K}= {\rm Im} (D^\dagger)$ define the kinematic 
   algebra with bracket defined by $D^\dagger$ together with invariant pairing, see
    \cite{Ben-Shahar:2024dju} for further details and explicit amplitude calculations. 
     Alternatively we can use the complex conjugate operator $\bar{D}^\dagger$ for the gauge fixing. 

 Next, for the purpose of constructing the double copy, let us double the odd 
  coordinates,  we have  $(\theta, \bar{\theta}, \zeta)$ with corresponding operators $D$ and $D^\dagger$ and we introduce another set
$(\theta', \bar{\theta}', \zeta')$ with the corresponding operators denoted as  $\D$, $\gen$. 
  Introduce a superfield ${\cal A} (x, \theta, \theta', \zeta, \zeta')$ of bi-degree $(1,1)$ (total degree $2$). If we choose two operators $D$ and $\gen$ then they define a dgLa. 
  If we restrict to ${\rm Im}(\gen)$ we obtain a dgLa with invariant pairing 
  \begin{equation}
    \langle a, b \rangle =    \int d^2x ~d^2 \theta~d^2 \theta'~ d\zeta~d\zeta'~ a~c
  \end{equation}
   with $b=\gen c$ and $a \in{\rm Im}(\gen)$. Thus on ${\cal A} (x, \theta, \theta', \zeta, \zeta') \in{\rm Im}(\gen)$ we define the BV symplectic structure
   \begin{equation}
       \omega_{BV} = \langle \delta \A, \delta \A \rangle~,  
   \end{equation}
    with the master action 
    \begin{equation}
       S_{D.C.} = \frac{1}{2} \langle \A, D \A \rangle + \frac{1}{6} \langle \A, \gen (\A^2) \rangle ~.
       \end{equation}
 Using the additional operator $\D$ we can rewrite above pairing and action 
  as the following non-local expressions,
\begin{equation}\label{2dYM-BVst-gen}
  \omega_{BV} =  \int d^2x ~d^2 \theta~d^2 \theta'~ d\zeta~d\zeta'~ \delta \A ~\frac{\D}{\Box} \delta \A
  \end{equation}
  and 
\begin{equation}\label{2d-YM-action-general}
 S_{D.C.} = \int d^2x ~d^2 \theta~d^2 \theta'~ d\zeta~d\zeta' \Big ( \frac{1}{2}{\cal A} \frac{\D D}{\Box} {\cal A} +  \frac{1}{6}{\cal A}^3 \Big )~,     
\end{equation}
  where we assume that ${\cal A}$ is in ${\rm Im} (\gen)$. 
  
 Due to the property (\ref{2DYM-sol-GF})  we have that ${\rm Im}(\gen)=\ker(\gen)$
 and we can parametrize this subspace explicitly in the following 
   fashion 
\begin{align}
 &{\cal A} (x, \theta, \theta', \zeta, \zeta') \nn \\ &= (\mathbf{A} (x, \theta, \theta') + \zeta {\cal I}'\mathbf\Phi (x, \theta, \theta') )  + \zeta'  2 \partial'^\dagger \Big ( {\cal I}' \mathbf{A} (x, \theta, \theta') - \zeta \mathbf\Phi (x, \theta, \theta') \Big ) \nn \\
 &= \gen \Big (-\zeta' {\cal I}'\mathbf{A}(x, \theta, \theta') + \zeta \zeta' \mathbf\Phi (x, \theta, \theta') ]  \Big  ) ~,
 \end{align}
 where we introduced superfields $\mathbf{A}$ of bi-degree $(1,1)$ (total degree $2$) and $\mathbf{\Phi}$ of bi-degree $(0,1)$ (total degree $1$). 
 If we take this parametrization of 
  ${\rm Im}(\gen)$ and integrate out $\zeta$ and $\zeta'$ 
 we get the BV symplectic  structure (\ref{2dYM-BVst-gen}) to become
\begin{equation}
  \omega_{BV} =  2 \int d^2x ~d^2 \theta~d^2 \theta'~ \delta \mathbf{A} \delta \mathbf{\Phi}
\end{equation}
  and the action (\ref{2d-YM-action-general}) will be   
 \begin{equation}\label{dc-2dYM-local}
      S_{D.C.} = \int d^2x ~d^2 \theta~d^2 \theta' \Big ( \mathbf{\Phi} d   \mathbf{A} -\frac{1}{2} \mathcal{I}'\mathbf{\Phi}~{\rm vol}~\mathbf{\Phi} + \frac{1}{2} \mathbf\Phi \{ \mathbf{A}, \mathbf{A}  \} \Big )~,
  \end{equation}
 where the Lie bracket $\{~,~\}$ is defined as follows
  \begin{equation}\label{2dYM-LA}
      \{\mathbf{A},\mathbf{A}\} = -2 {\partial'}^\dagger (\mathbf{A}\mathbf{A}) - 4\mathcal{I}'(\mathbf{A}\mathcal{I}'{\partial'}^\dagger \mathbf{A}) ~. 
  \end{equation}
This Lie algebra corresponds to the kinematic algebra for 2d YM has been discussed explicitly in in \cite{Ben-Shahar:2024dju}.
Note that by removing the $\mathbf{\Phi}^2$ from equation \eqref{dc-2dYM-local} one obtains the double copy of 2d YM with 2d BF theory, where the fields are constrained to be int he image of the gauge-fixing operator from 2d YM side of the double copy.
 Similarly to the previous section, the double copy (\ref{dc-2dYM-local}) resembles a deformed BF theory with gauge symmetry given by the infinite dimensional Lie algebra 
    (\ref{2dYM-LA}), however this interpretation is not quite correct, as once again the space on which the kinematic algebra acts is the spacetime of the fields. Let us write this theory in terms of physical fields (fields of ghost degree $0$), denoting $\theta$ as $\theta^\mu$ with Greek indices and 
     $\theta'$ as $\theta'^a$ with Latin indices and remember that $d= \theta^\mu \frac{\partial}{\partial x^\mu}$. The physical components of
      $\mathbf{A}$ are,
\begin{equation}
      \mathbf{A}|_{\rm phys} = e_{a\mu} \theta'^a \theta^\mu + \phi ~\epsilon_{ab} \theta'^a \theta'^b  + e~ \epsilon_{\mu\nu} \theta^\mu \theta^\nu 
\end{equation}
      and the physical components of $\mathbf{\Phi}$ are 
\begin{equation}
       \mathbf{\Phi}|_{\rm phys} = \epsilon_{ab} \phi^b \theta'^a + \omega_\mu \theta^\mu~. 
\end{equation}
 The physical part of the action (\ref{dc-2dYM-local}) is given by the following expression 
 \begin{align}\label{e:2d-dc-physical}
 S_{D.C.}|_{\rm phys} = \int \Big ( &
\phi_{z'}de_{\bar{z}'}
+\phi_{\bar{z}'}de_{z'} 
-\phi d w
-\frac{1}{2}\phi_{z'}\phi_{\bar{z}'}
-\phi_{z'}\phi \partial_{\bar{z}'}e\\
&+\phi \partial_{\bar{z}'}w e_{z'} 
-2\phi_{z'} e_{\bar{z}'}\partial_{\bar{z}'} e_{z'}
+\phi_{z'}e_{z'}\partial_{\bar{z}'}e_{\bar{z}'}
+\phi_{\bar{z}'}e_{z'}\partial_{\bar{z}'}e_{{z}'}
 \Big ) \nonumber
 \end{align}
after also rescaling $\mathbf{A}\to \mathbf{A}/4$ and $\mathbf{\Phi}\to -2i\mathbf{\Phi}$ to clean up numerical factors, and $e_a$, $\omega$ are interpreted as one-forms while $\phi^a$, $\phi$ as zero-forms. The prime on the free indices only serves to remind us that these indices came from the right-hand-side of the double copy, and there is no distinction between primed and unprimed coordinates at this stage.

 We can gauge fix the action (\ref{dc-2dYM-local}) by restricting the fields to ${\rm Im}(\bar{D}^\dagger)$, which, as before, can be resolved algebraically. With this gauge choice the action becomes,
 \begin{equation}
      S_{D.C.} = \int d^2x ~d^2 \theta~d^2 \theta' \Big ( \mathbf{A} \Box   \mathbf{A} + \frac{1}{2} \mathbf{A} \{\{ \mathbf{A}, \mathbf{A} \} \} \Big )~,
      \end{equation}
   where $\mathbf{A} (x, \theta, \theta')$ is an unrestricted superfield and the double bracket $\{\{~,~\}\}$ (which is not a Lie bracket by itself) denotes the tensor product of the Lie bracket coming 
    from $\gen$ and from $D^\dagger$ (or $\bar{D}^\dagger$). If we chose $\gen$ on one side and $\bar{D}^\dagger$
      on the other side then the resulting amplitudes will be real in Euclidean signature.

\section{ A general class of solutions of CME}\label{s:formal}

  In this section we summarize the algebraic data that have been used to construct the double copy solution of the CME. These data define a general class of solutions of the CME, whose examples do not all come from the double copy construction. Remarkably, in a broad sense they resemble gravity theories. 

Consider a bigraded manifold $\M$; we denote with $C(\M)$ its functions 
 and with $(d_L,d_R)$ the two degrees. We call $d_{L+R}=d_L+d_R$. We require the following data
\begin{itemize}
 \item[$i$)] an integral $\int~d\mu:C(\M)\rightarrow \Re$ of degrees $(-3,-3)$;
 \item[$ii$)] a vector field $D\in\Vect(\M)$ of degrees $(1,0)$; 
 \item[$iii$)] a second-order operator $\gen$ of degrees $(0,-1)$
\end{itemize}
such that
\begin{equation}
\label{general_requirements_1}
D^2=0~,~~~~(\gen)^2 =0~,~~~~D\gen + \gen D =0
\end{equation}
and the following compatibility conditions between the integration and 
$D, \gen$  
\begin{equation}
 \label{general_requirements_2}
 \int D(a)b = (-1)^{|a|+1}\int a D(b)\,,\;\;
 \int \gen(a)b = (-1)^{|a|}\int a \gen(b)
\end{equation}
for each $a,b\in C(\M)$ and $|a|=d_{L+R}(a)$. 

Data ($i-iii$) define a solution of the CME. Let us define the dgLa $\dgLa=({\rm Im}(\gen), D, $\{~,~\}$, \langle~,~\rangle)$ with degree $\deg=d_{L+R}-1$,
bracket $\{~,~\}$ defined as 
 \be
 \{a,b\} = \gen (ab)~,~~~~~a,b \in {\cal K}\;
 \ee

and invariant pairing  $\langle~,~\rangle$
defined as
\begin{equation}\label{integ-gen}
\langle a,b \rangle = \int d\mu~a\xi~,
\end{equation}
where $b=\gen\xi$. The pairing has the following symmetry property
\begin{equation}
    \langle a,b \rangle = (-1)^{\deg a \deg b}\langle b, a\rangle =
 (-1)^{(d_{L+R} (a)+1) (d_{L+R} (b)+1)}   \langle b, a \rangle~.
\end{equation}  
It is clear that the pairing has $d_{L+R}=-5$ and $\deg = -3$. The superfield $\A$ has total degree ${\rm tot} = d_{L+R}+\gh = \deg + \gh +1 =2$. As usual the  
symplectic form is
\begin{equation}\label{bv_symplectic}
    \omega = \langle \delta \A,\delta \A\rangle  \ , 
\end{equation}
and the action 
\begin{equation}\label{bv_action}
    S = \frac{1}{2}\langle \A , D \A\rangle + \frac{1}{6}\langle \A,\{\A,\A\}\rangle \ ,
\end{equation}
solves CME. Examples of data ($i-iii$) are given by $\so(3,\Re)$ CS discussed in subsection \ref{ss:so(3)} and $JT$-gravity discussed in subsection \ref{ss:JT}.

We  can introduce the following additional data:
\begin{itemize}
 \item[$iv$)] a vector field $\D\in\Vect(\M)$ of degree $(0,1)$ such that
\end{itemize}
  \begin{equation}
    \D^2=0~,~~~~~  \D D + D \D=0
  \end{equation}
 and 
\begin{equation}
    \int \D(a)b = (-1)^{|a|+1}\int a \D(b)
\end{equation}
  with $|a|=d_{L+R}(a)$.
  Moreover  we can define the generalized Laplace operator as
 \begin{equation}
\square \equiv  [\D,\gen]~.
\end{equation}
In this case the pairing $\langle~,~\rangle$ on $\dgLa$ can be written as
\begin{equation}
\langle a,b\rangle = \int d\mu~ a \frac{1}{\square} \D b~,
\end{equation}
 provided that we can invert $\square$. 
  In this case  the BV symplectic structure (\ref{bv_symplectic})  and the solution of CME (\ref{bv_action}) 
  can be expressed as the integral of a density that is a (non local) functional of the fields
\begin{equation}\label{BVsymp-non-local}
\omega = \frac{1}{2} \int d\mu~ \delta\A \frac{1}{\square} \D   \delta \A
\end{equation}
  and  
\begin{equation}\label{dgLA_action-non-local}
S_{BV}(\A) = \int d\mu \Big  (\frac{1}{2} \A \frac{1}{\square} \D D \A
+ \frac{1}{6}  \A^3 \Big ) ~,
\end{equation}  
 for $\A \in \dgLa$. To make everything well-defined we need to take care of the zero modes $\ker\Box\cap{\rm Im} \gen$ and this depends on the details of the theory, so we do not address this problem here. Finally we can introduce 
\begin{itemize}
 \item[$v$)] a second-order operator $D^\dagger$, of degree $(-1,0)$ 
\end{itemize}
satisfying
\begin{equation}
    \square = [D,D^\dagger]~,~~~(D^\dagger)^2=0~,~~~D^\dagger \gen + \gen D^\dagger=0~,~~~D^\dagger \D + \D D^\dagger=0
\end{equation}
 and 
\begin{equation}
\int D^\dagger(a)b = (-1)^{|a|}\int a D^\dagger(b)~.
\end{equation}
 Then ${\rm Im} (D^\dagger)$ is a Lagrangian submanifold for (\ref{BVsymp-non-local}) and so defines a gauge fixing. Note that we take the Laplacians from both sides of the double copy construction to agree, meaning $[D,D^\dagger]=[\D,\gen]=\Box$, see for example ref. \cite{Bonezzi:2022bse} for further discussions of this constraint.

 If conditions ($i-v$) are satisfied, there is a symmetry in the algebraic data so that the role of the operators $(D, \gen)$ defining the solution of CME can be exchanged with those $(\D, D^\dagger)$ defining the gauge fixing. We then have two possible solutions of the CME that share the same gauge-fixed action. In particular, this implies that the gauge-fixed action ($S_{BV}({\cal A})$ restricted to ${\cal A} \in {\rm Im} (D^\dagger) \cap {\rm Im} (\gen)$) enjoys two BRST symmetries that are the residual symmetry of the BV vector fields restricted on the Lagrangian submanifold
\begin{align}\label{double_BRST}
    Q {\cal A} = \Big (  D \A + \frac{1}{2} \gen \A^2 \Big )\Big |_{{\rm Im}(D^\dagger)}~ , \ \ \  \ \ \ \mathcal{Q} \A = \Big ( \D \A - \frac{1}{2} D^\dagger \A^2\Big )\Big|_{{\rm Im}(\gen)} ~ .
\end{align}
These two BRST transformations obey 
$Q^2=0$, $\mathcal{Q}^2=0$ and $Q \mathcal{Q} + \mathcal{Q} Q=0$ on shell. Indeed in Appendix \ref{a:BRST} we will see that there is actually a family of such solutions of CME, parametrized by $SL(2)$.

We are going to see in the next sections that there are other interesting examples of these data that do not come from the double copy construction.

\section{Kodaira-Spencer gravity revisited}\label{s:KS}

  In this section we want to illustrate that the BV description of the Kodaira-Spencer theory fits the formal framework described in section \ref{s:formal}.
The Kodaira-Spencer gravity was introduced in \cite{Bershadsky:1993cx} together 
 with its BV description. The Kodaira-Spencer gravity has attracted considerable 
  attention during last 30 years, here we concentrate only on specific BV aspects of this theory and its generalizations. Previously the relation between 
   the CS double copy and the Kodaira-Spencer gravity has been already remarked in \cite{Bonezzi:2024dlv} 

\subsection{6D Kodaira-Spencer gravity}

Consider CY3 fold $M_6$  
equipped with the holomorphic
volume form $\Omega$. 
 For the sake of simplicity let us work in coordinates 
 when $\Omega$ is constant 
\be
 \Omega = \frac{1}{3!} \epsilon_{i j k} dx^i \wedge dx^j \wedge dz^k = dx^1 \wedge dx^2 \wedge dx^3~,
\ee
 where $(x^1, x^2, x^3)$ are holomorphic coordinates and $(x^{\bar{1}}, x^{\bar{2}}, x^{\bar{3}})$ are anti - holomorphic coordinates on $M_6$. All formulas below have straightforward generalization to any complex coordinates when $\Omega$ is not constant. 

 The bi-graded manifold $\M = (T^{(0,1)}[1]\oplus T^{*(1,0)}[1]) M_6 $
  with  odd coordinates $\theta^{\bar{i}}$ with $\dl=1$ (transform as $d x^{\bar{i}}$) and $\theta_i$  with $\dr=1$ (transform as $\partial_i$). 
  Thus the functions of bidegree $(k,p)$ are identified with 
   $(0,k)$-forms valued in $\wedge^p T^{(1,0)}$, {\it i.e.} $C^{(k,p)}(\M) = \Omega^{(0,k)} (M, \wedge^p T^{(1,0)})$. 
   Let us introduce the following 
   operators defined on this space 
   \be\label{KS-def-D1}
      D = \theta^{\bar{i}} \frac{\partial}{\partial x^{\bar{i}}}~,~~~~
      \gen = \frac{\partial^2}{\partial \theta_i \partial x^i}~,
   \ee
    where the first operator $D$ is vector field of bi-degree $(1,0)$ such that $D^2=0$  and which is canonically defined for any complex manifold. The operator $\gen$ is second order of bi-degree $(0,-1)$ and it  requires a holomorphic volume form to be defined globally. 
     Obviously these two operators anti-commute, 
     $\{ D , \gen \}=0$. Next we define the integration of bi-degree $(-3,-3)$
     \begin{equation}
         \int d\mu = \int  d^6 x~d^3\theta~d^3\bar{\theta}~,
     \end{equation}
 which is non-canonical and it requires the holomorphic volume form. 
  The integration is compatible with $D$ and $\gen$ as in (\ref{general_requirements_2}).
  On $C(\M)$ the operator $\gen$ generates the bracket on the superfields 
      \be\label{algebra-KS}
       \{ \A_1, \A_2 \} = \gen (\A_1 \A_2 ) - \gen (\A_1) \A_2 - (-1)^{|\A_1|} \A_1 \gen(\A_2)~,
      \ee
    which descends to the bracket on ${\cal K} = {\rm Im} (\gen)$. On 
 ${\rm Im} (\gen)$ we can define the pairing as in (\ref{integ-gen})
  which is non-degenerate.
    Moreover this pairing is invariant with respect to the bracket $\{~,~\}$. 

  We introduce a superfield $\A (x, \theta, \bar{\theta})$ of bi-degree $(1,1)$ and its components can be identified with $\Omega^{(0,k)} (M, \wedge^p T^{(1,0)})$. We now have the data $(i-iii)$ of section \ref{s:formal} so that we can define the BV action for KS gravity as the follows 
    \be
      S_{KS} = \frac{1}{2} \langle \A, D \A \rangle + \frac{1}{6} \langle \A, \{ \A, \A \} \rangle~
    \ee
     with the BV symplectic structure 
     \be
      \omega_{BV} = \langle \delta \A, \delta \A \rangle~. 
     \ee
      Remember that we define both BV symplectic structure and $S_{KS}$ over the space $\A \in {\rm Im} (\gen)$. The above construction requires 
       only holomorphic volume form. The equations of motion for the above action are given by   
       \be\label{deform-KS-basic}
        D\A + \frac{1}{2} \{\A , \A\} = 
        D \A + \frac{1}{2} \gen (\A^2)=0~,
       \ee
  which is the Maurer-Cartan equation which we discuss below. 
  We can extend our formulas from ${\rm Im} (\gen)$ to $\ker (\gen)$
       \be
        \A = \gen \xi + a~,
       \ee
 such that $\gen \A=0$. The algebra (\ref{algebra-KS}) descends to $\ker (\gen)$.
  While the pairing can be extended to whole $\ker (\gen)$ but it becomes degenerate, $\langle a_1, a_2 \rangle =0$. So either we deal with the BV theory 
   with degenerate BV symplectic structure (we still can define brackets and solve 
    master equation) or we consider the standard BV theory expanded around given $a$.

 If we assume that CY manifold $M_6$ is equipped with compatible K\"ahler metric
  then more operations can be defined. Using K\"ahler metric we define two additional operators
      \be\label{KS-def-D2}
          \D = \theta_i g^{i \bar{j}} \frac{\partial}{\partial x^{\bar{j}}}~,~~~~D^\dagger = g^{i \bar{j}} \frac{\partial^2}{\partial \theta^{\bar{j}} \partial x^i}
      \ee
  where for clarity  we use the flat metric (otherwise we 
 need to covariantize the derivatives in $x$). Here $\D$ is differential of bi-degree $(0,1)$ and $D^\dagger$ is second order operator of bi-degree $(-1,0)$.
 The operators satisfy
       \be
        \Box = \{ D, D^\dagger \} = \{ \D, \gen \}~,
       \ee
 where $\Box$ is generalised Laplace operator.  All these operators satisfy the properties described in the subsection \ref{s:formal}. 
 Thus on a K\"ahler manifold the BV action can be rewritten in the following non-local form  
   \be
      S_{KS} = \int d^6x~d^3 \theta~ d^3\bar{\theta} \Big ( \frac{1}{2} \A \frac{\D D }{\Box} \A + \frac{1}{6} 
      \A^3 \Big )~,
     \ee
 where we always assume that $\A \in {\rm Im} (\gen)$. 

Next let us discuss the geometrical meaning of the equation of motion (\ref{deform-KS-basic}) restricted to the physical fields.    
 Let us restrict to the physical part of $\A$ 
 \be
  \A |_{\rm phys} = A^i_{\bar{j}} \theta_i \theta^{\bar{j}} + B_{\bar{i}\bar{j}} \theta^{\bar{i}} \theta^{{\bar{j}}} + C^{ij} \theta_i \theta_j~,
 \ee
 which correspond to element in $\wedge^2 L$ with $L= T^{(0,1)} \oplus T^{*(1,0)}$.  Since $\gen \A=0$ we have  $\partial_i A^i_{\bar{j}} =0$
  and $\partial_i  C^{ij}  =0$. The last condition actually implies 
  automatically 
  \begin{equation}
    C^{ij} \partial_j C^{kl} + {\rm cyclic}(i,k,l)=0~,  
  \end{equation}
  and this is true only in 6D case. 
  The equation (\ref{deform-KS-basic}) for physical fields become  
  \bea
 &&  \partial_{\bar{i}} B_{\bar{k}\bar{j}} + A^l_{\bar{i}} \partial_i B_{\bar{k} \bar{j}} + {\rm cyclic}(\bar{i},\bar{k}, \bar{j}) =0~,\\
 &&  \partial_{[\bar{s}} A_{\bar{p}]}^l + A^i_{[\bar{s}} \partial_i A^l_{\bar{p}]} + 4 C^{li} \partial_i B_{\bar{s}\bar{p}}=0~,\\
 && \partial_{\bar{k}} C^{ij}  + A^s_{\bar{k}} \partial_s C^{ij} + C^{js} \partial_s A^i_{\bar{k}} - C^{is} \partial_s A^j_{\bar{k}} =0
  \eea
If we set $C^{ij}$ and $B_{\bar{i}\bar{j}}$ to 
  zero then we get the standard story for the deformations of complex structure
   with the additional constraint $\partial_i A^i_{\bar{j}}=0$. If we keep all fields then these equations correspond to deformation of  generalized complex structure, see Gualtieri's thesis  \cite{Gualtieri:2003dx} (previously these generalized deformations were discussed by Kontsevich and Barannikov \cite{MR1609624}).
  In the above equations $C^{ij}$ is understood as unimodular holomorphic Poisson structure and $B_{\bar{i}\bar{j}}$ are closed anti-holomorphic 
   $b$ form.  It is important to stress that BV formulation of KS gravity within our
    framework requires to include the generalized deformations as part of the picture. 

  The most natural language for previous discussion is in terms of generalized complex, generalized K\"ahler and generalized Calabi-Yau geometries.  In this language the holomorphic volume form $\Omega$ is pure spinor for corresponding complex structure (understood as generalized complex structures) and it provides the map to differential forms 
  \be
    \A ~\rightarrow~ \A \cdot \Omega~,
  \ee
   where we contract the holomorphic vector indices with holomorphic form and
    get the differential form.
   Under this map we have the following relation between different operators
   \bea
  && ( D \A) \cdot \Omega = \bar{\partial} (\A\cdot \Omega)~,\\
   && (\gen \A ) \cdot \Omega = \partial (\A \cdot \Omega)~, \\
     && (D^\dagger \A) \cdot \Omega = \bar{\partial}^\dagger (\A \cdot \Omega)~,\\
   && (\D \A ) \cdot \Omega = \partial^\dagger (\A \cdot \Omega)~, 
   \eea
   where $\partial$, $\bar{\partial}$, $\partial^\dagger$, $\bar{\partial}^\dagger$
   are standard differentials associated with K\"ahler manifold. 
   For more details see
 \cite{Gualtieri:2003dx}. It is obvious that one can extended this construction
  of the Kodaira-Spencer gravity to other generalized Calabi-Yau 6 dimensional manifolds if one uses the language of the generalized geometry. This is outside of the scope of the present paper and we leave it for the future. 

\subsection{Kodaira-Spencer gravity in other dimensions}

 In previous subsection we have described how to relate the standard 
 Kodaira-Spencer 6D theory to the formal construction inspired by the BV double copy description. Actually this construction can be easily applied to other theories. As an illustration, let us briefly discuss 8-dimensional analog of the Kodaira-Spencer theory defined for a CY 4-fold $M_8$. 

Let us assume that $M_8$ is equipped with a holomorphic volume form and K\"ahler metric. Let us consider the bi-graded manifold  ${\cal M}\equiv(T^{(0,1)}[1]\oplus T^{*(1,0)}[1])) M_8 $ with 
odd coordinates $\theta^{\bar{i}}$ with $\dl=1$  and $\theta_i$  with $\dr=1$. The function on this bi-graded manifold are identified with 
   $(0,k)$-forms values in $\wedge^p T^{(1,0)}$, $C(\M) = \Omega^{(0,k)} (M, \wedge^p T^{(1,0)})$. Following previous subsection we can define the same operations:  $D$ and $\gen$ as in (\ref{KS-def-D1}) and  
 $D^\dagger$ and $\D$ as in (\ref{KS-def-D2}). The main problem in 8D is that 
  the integration $d^8x~d^4 \theta~d^4\bar{\theta}$ has wrong bi-degree $(-4,-4)$. The way out is to introduce two additional odd coordinates:
  odd $\zeta$ of bi-degree $(-1,0)$ and odd $\bar{\zeta}$
 of bi-degree $(0,-1)$. Thus the measure 
\begin{equation}
  \int~d^8x~d^4\theta~d^4\bar{\theta}~d \bar{\zeta}~d \zeta   
\end{equation}
 will be of bi-degree $(-3,-3)$. Thus we need to introduce the superfield
 $\A (\theta, \bar{\theta}, \zeta, \bar{\zeta})$ of bi-degree $(1,1)$
 which has the following expansion
\begin{equation}
    \A (\theta, \bar{\theta}, \zeta, \bar{\zeta}) = \mathbf{A} (\theta, \bar{\theta}) + \zeta \mathbf{A}_\zeta (\theta, \bar{\theta})
 +  \bar{\zeta} \mathbf{A}_{\bar{\zeta}} (\theta, \bar{\theta})
  +  \zeta \bar{\zeta} \mathbf{A}_{\zeta\bar{\zeta}}(\theta, \bar{\theta})
\end{equation}
 and thus we identify the components of this superfield with four copies of 
 $\Omega^{(0,k)} (M, \wedge^p T^{(1,0)})$. Here $\mathbf{A}$ is of bi-degree $(1,1)$, $\mathbf{A}_\zeta$ is of bi-degree $(2,1)$, $\mathbf{A}_{\bar{\zeta}}$ is of bi-degree $(1,2)$ and  $\mathbf{A}_{\zeta\bar{\zeta}}$ is of bi-degree $(2,2)$.  
   If we assume that ${\cal A} \in {\rm Im}(\gen)$ then we can write the following solution of the master equation
  \begin{equation}\label{8D-KS-action}
     S_{KS} =   \int~d^8x~d^4\theta~d^4\bar{\theta}~d \bar{\zeta}~d \zeta 
      \Big (  \frac{1}{2}\A \frac{D\D}{\Box} \A + \frac{1}{6} \A^3 \Big )
  \end{equation}
  with respect to BV symplectic structure 
\begin{equation}
     \omega_{BV} = \int ~d^8x~d^4\theta~d^4\bar{\theta}~d \bar{\zeta}~d \zeta ~
     \delta \A ~\frac{\D}{\Box} \delta \A~.
\end{equation} 
 This fits completely within our formal framework described in section \ref{s:formal}
  and all statements from there can be extended to this 8D Kodaira-Spencer theory. This construction admits a straightforward generalization to other even dimensions, but the benefit of double-copying 4d BF theory is that, as mentioned before, the condition ${\cal A} \in {\rm Im}(\gen)$ results in a local action when $\gen = \bar{\partial}^\dagger +\lambda \zeta$.
 Without explicitly solving the ${\cal A} \in {\rm Im}(\gen)$ condition though, if we perform the integration $d\bar{\zeta}~d\zeta$ in the action (\ref{8D-KS-action}) we get the following expression 
\begin{equation}
 S_{KS} = \int d^8x~ d^4\theta~ d^4\theta~ \Big ( \mathbf{A}_{\zeta\bar{\zeta}}
 \frac{ D\D}{\Box} \mathbf{A} - \mathbf{A}_\zeta  \frac{ D\D}{\Box} \mathbf{A}_{\bar{\zeta}} + \frac{1}{2} \mathbf{A}_{\zeta\bar{\zeta}} \mathbf{A}^2 - \mathbf{A} \mathbf{A}_\zeta \mathbf{A}_{\bar{\zeta}} \Big )
\end{equation}
 and similar for the BV symplectic structure
 \begin{equation}
      \omega_{BV} = 2 \int ~d^8x~d^4\theta~d^4\bar{\theta}~ \Big ( \delta \mathbf{A} ~\frac{\D}{\Box} \delta \mathbf{A}_{\zeta\bar{\zeta}} + 
       \delta \mathbf{A}_{\zeta} ~\frac{\D}{\Box} \delta \mathbf{A}_{\bar{\zeta}} \Big )~.
 \end{equation}
 In this theory the deformations of generalized complex structure are encoded in the physical sector of the superfield $\mathbf{A}$, but there are more physical field in this theory and more equations which require further geometrical investigation. 

Similar non-local actions can be constructed from BF theories in other dimensions, and the equations of motion in the physical sector will contain similar equations of motion for a generalized complex structure as in the previous subsection.  
It is also possible to double copy $YM^2$, $BF^2$, and $YM\otimes BF$, starting with theories in 2d and arriving at actions in 4d. These double copies basically take the form of the actions already presented \eqref{dc-2dYM-local}, \eqref{e:2d-dc-physical}, but in the latter action the $z$ and $z'$ coordinates are conjugate and the primed indices should be raised. It would be interesting to investigate the geometrical details of these theories in future work.

\subsection{Theory of K\"ahler gravity}

One of the simplest examples of the algebraic data described in section \ref{s:formal} is 6D K\"ahler manifold. Consider the bi-graded supermanifold $\M = T[1] \Sigma_6$ where  $\Sigma_6$ is a six dimensional K\"ahler manifold, where $d_L(\theta^i)= d_R (\theta^{\bar{i}})=1$ and $d_L(\theta^{\bar{i}})=d_R (\theta^i)=0$ in complex coordinates. We have the collection of the operators $\partial$, $\bar{\partial}$, $\partial^\dagger$ and $\bar{\partial}^\dagger$  and  
    $\M$ is equipped with  canonical integration $\int d^6 x~d^6\theta$ of bi-degree $(-3,-3)$. 
     All these operations satisfy the relations listed in subsection \ref{s:formal}. Let us consider
      instead linear combination of  the  operators: 
  $D=d$, $D^\dagger= d^\dagger$, $\D = d^c$ and $\gen=d^{c\dagger}$,
   where $d^c = i (\bar{\partial} - \partial )$. These operators respect total degree. We can define the superfield $\A$ of total degree $2$
   and write the following action 
\begin{equation}
  S_{BV} = \int d^6x~d^6\theta~ \Big ( \frac{1}{2} \A \frac{d d^c}{\Box} \A + \frac{1}{6}\A^3 \Big ) 
\end{equation}
 with $\A \in {\rm Im}(d^{c\dagger})$. This action satisfies the CME with respect to the BV symplectic 
  structure 
  \begin{equation}
      \omega_{BV} = \int d^6x~d^6\theta~ \delta \A \frac{d^c}{\Box} \delta \A
  \end{equation}
  This theory is the BV formulation of the 6D theory of K\"ahler gravity which has been studied in \cite{Bershadsky:1994sr}. As mentioned in the previous subsection, similar theories can also be constructed by using two copies of BF theory in various dimensions, and in 2d in particular the data from 2d YM can also be used.

\section{Summary and open questions}\label{s:end}

In \cite{Ben-Shahar:2021zww}, a gauge-fixed action for the double copy of Chern-Simons theory was proposed. The goal of this work is to understand that construction within the BV formalism and to generalize it to other theories. This study is a natural continuation of our previous paper \cite{Ben-Shahar:2024dju}, where we developed a framework for gauge theories exhibiting off-shell color-kinematics duality. 

Here, we propose a BV formulation for double copy theories that possess off-shell color-kinematics duality and formalize the underlying construction. 
These constructions can be used to write gravity-like actions in general even dimensions, but a few special examples stand out for having local actions. These are possible when either 4d BF theory or 2d YM theory are used.

The construction described here from the BV formalism offers two useful features. First, it provides a natural framing of the result in terms of a Maurer-Cartan equation, which we interpret as encoding a deformation problem, as illustrated in our examples.
Second, these theories exhibit an intriguing feature: their gauge-fixed actions possess two anti-commuting BRST symmetries. This arises because the same gauge-fixed action can be obtained by applying different gauge-fixing conditions to two distinct BV theories. We suggest that this dual BRST structure may be viewed as a defining property of these double copy theories.
 
Many open questions remain within our proposed framework. A deeper analysis of the relevant theories is needed, including a more thorough investigation of the associated deformation equations and their potential gravitational interpretations. It would also be valuable to connect these issues to explicit off-shell computations of observables in the corresponding gravity theories.

In this work, we identified a gauge theory - namely, 4D BF theory - that admits two distinct kinematic algebras, which are not evidently related to one another. Although these two algebras lead to the same physical theory (as they correspond to different gauge fixings of the same gauge theory), the double-copy constructions they give rise to look superficially quite different, one being non-local and containing constrained one- and two-form fields, while the other is local and has unconstrained two-form fields. This situation is puzzling, and we conjecture that the two kinematic algebras and the corresponding double copy theories should be related, perhaps through a (potentially non-local) field redefinition.

One of the major open challenges is to determine whether a double copy theory can be constructed without relying on off-shell color-kinematics duality, see for example ref. \cite{Gonzalez:2021bes}. Concretely, this would correspond to the case where the operator 
$\gen$ is not of second order. In such a scenario, can one still construct a BV action that satisfies the master equation?
It is possible that the resulting action would involve higher-order terms and fall outside the scope of the formal Chern-Simons framework. We hope to revisit this important question in future work.

\subsubsection*{Acknowledgements}

We are grateful to  Henrik Johansson and Roberto Bonezzi for enlightening discussions.
The research of M.\ B.\ S.\ is supported by the Knut and Alice Wallenberg Foundation
(grant KAW 2023.0490).
 The research of M.\ Z.\ 
   is  supported by the the Swedish Research Council excellence center grant ``Geometry and Physics'' 2022-06593. This
 work is also supported by the Swedish Research Council under grant no. 2021-06594
while all authors  were in residence at Institut Mittag-Leﬄer in Djursholm, Sweden during January-April of 2025.

\appendix

\section{BV formulation of Chern-Simons theory}\label{s:CS}

In this appendix we review the standard construction of the Chern Simons solution of the Classical Master Equation (CME) within the BV formalism. The aim is to highlight the features that will allow the  generalizations discussed in the next sections, including the double copy theories. The key ingredient needed to construct a solution to the CME is a cyclic dgLa.

Let $\Sigma_3$ denote a closed $3$-manifold and  $\mathfrak{g}$ denote a quadratic Lie algebra. Consider the differential graded Lie algebra (dgLa) $\Omega(\Sigma_3)\otimes\g$ with differential given by the de Rham differential $d$ and the bracket $[~,~]$ defined by the wedge product on forms and the Lie bracket of $\g$. We denote $\deg$ the form degree. If ${\rm Tr}$ denotes an invariant non-degenerate  pairing on $\mathfrak g$, the pairing on $\Omega (\Sigma_3) \otimes\g$  defined as  
\begin{equation}\label{cs_dgLa_pairing}
\langle \omega\otimes a,\nu\otimes b\rangle = \int_{\Sigma_3}\omega\wedge\nu \ {\rm Tr}(ab)
\end{equation}
is nondegenerate and of degree $-3$. Let us assign the ghost degree by the rule
$$
1 = \gh + \deg\;.
$$
Let $\A\in\Omega(\Sigma_3)\otimes\g$ be decomposed in the form degree as
$$
\A = A_0+A_1+A_2+A_3
$$
where $\gh (A_r)=1-r$. The symplectic form 
\begin{equation}\label{BV_sympl_form}
\omega = \frac{1}{2}\langle \delta\A,\delta\A\rangle = \int_{\Sigma_3}\Tr(\delta A_0\delta A_3) + \frac{1}{2}\Tr(\delta A_1\delta A_2)
\end{equation}
has ghost degree $-1$. The action
\begin{equation}\label{CS_BV_action}
S_{CS}(\A) = \frac{1}{2}\langle \A,d\A\rangle +\frac{1}{6}\langle\A,[\A,\A]\rangle
\end{equation}
solves the CME
\begin{equation}
 \label{CME}
 \{S_{CS},S_{CS}\}=0
\end{equation}
where $\{~,~\}$ is the degree $1$ bracket defined by (\ref{BV_sympl_form}).

It is important to notice that this construction of a solution of the CME (\ref{CME}) depends only on the dgLa data of $\Omega(\Sigma_3)\otimes\g$. Thus we can formalize the above properties as the following ones: 
let $(\dgLa,D,[~,~])$ be a dgLa endowed with a degree $-3$ non degenerate symmetric pairing $\langle~,~\rangle$. Denoting with $\deg$ the degree, the pairing satisfies for each $a,b,c\in{\cal L}$
\begin{equation}
\label{dgLa_pairing}
\langle a,b\rangle = (-1)^{\deg a\deg b} \langle b,a\rangle\,,\;\;
\langle [a,b],c\rangle =\langle a,[b,c]\rangle
\end{equation}
and
\be
\langle Da,b\rangle = (-1)^{1+\deg a}\langle a,Db\rangle~.
\ee
 We assume that $\dgLa$ is bounded, {\it i.e.} there exists $r\geq 0$ such that
$$
\dgLa= \dgLa_{-r+3} \oplus \dgLa_{-r+4}\oplus\ldots \oplus\dgLa_r\;,
$$
where $\dgLa_i$ denotes the space of homogeneous elements of $\deg=i$. Clearly, the Chern-Simons dgLa is $\dgLa=\Omega(\Sigma_3)\otimes\g$ and $r=3$. Accordingly the superfield $\A \in \dgLa$ decomposes in homogeneous components as
$$\A= A_{-r+3} + A_{-r+4}+\ldots + A_r\;.$$
As before we define the ghost degree by $1 = \gh + \deg$. The symplectic form
\begin{equation}
\label{dgLA_symplectic_form}
\omega = \frac{1}{2}\langle \delta\A,\delta \A\rangle
\end{equation}
has ghost degree $-1$. The action
\begin{equation}\label{dgLA_action}
S_{BV}(\A) = \frac{1}{2}\langle\A, D\A\rangle + \frac{1}{6} \langle\A,[\A,\A]\rangle
\end{equation}
has ghost degree $0$ and solves the CME. In the following Sections, we will repeatedly use this construction with different choices of the above dgLa data.

\section{Deformation of Courant brackets}\label{a:deformations}

 This appendix provides the geometrical derivation of the equation 
  (\ref{eqom_MC_A-0})-(\ref{eqom_MC_A-1}). The purpose of this appendix is to illustrate a geometric interpretation of the equations of motion of the double copy of CS theory. 

 The main idea is that the Lie bracket of vector fields can be extended to the Courant bracket of vector fields plus one forms.
  The Courant bracket is not a Lie bracket but it plays a central role in generalized geometry. We want to start from 
  the flat vector fields and differential forms $(\partial_\mu, dx^\nu)$ with zero Courant brackets among them and we want to 
   deform the vector fields and one forms in such a way that the brackets stay zero. 

 The calculation can be organized using the language of the graded manifolds to encode the structure of the Courant algebroid (vector bundle with Courant bracket). For the details of this language we refer to refs.
  \cite{Roytenberg:1999mny,Roytenberg:2002nu}.

  The standard Courdant algebroid on $T^* \oplus T$ can be encoded 
   in terms of graded symplectic manifold $T^*[2]T[1]$ with sympelctic form of degree $2$
    \be
   \omega = dp_\mu \wedge dx^\mu + d v^\mu \wedge d q_\mu~,
  \ee
  where the coordinates $(x,p,v,q)$ are of degree $0,1,1,2$ respectively. Also the structure is encoded in terms of a
   Hamiltonian of degree $3$, in the flat case we can choose $\Theta = p_\mu v^\mu$. Now consider a symplectomorphism
 to the coordinates $(\tilde{x}, \tilde{p}, \tilde{v}, \tilde{q})$ with symplectic structure 
   \be
   \omega = d\tilde{p}_\mu \wedge d\tilde{x}^\mu + d \tilde{v}^\mu \wedge d \tilde{q}_\mu~.
  \ee
   Let us parametrize the symplectormorphism as follows
   \bea
     &&  x^\mu = \tilde{x}^\mu~,\nn \\
      &&  \tilde{v}^\mu = E^\mu_\nu v^\nu + 2 C^{\mu\nu} \tilde{q}_\nu~,  \label{sympl_transf_1}\\
      &&  q_\mu = E^\nu_{\mu}  \tilde{q}_\nu + 2 B_{\mu\nu} v^\nu~,\nn \\
      &&  p_\mu = \tilde{p}_\mu   - \partial_\mu A^\rho_{\nu}  v^\nu \tilde{q}_\rho -\partial_\mu C^{\rho\nu} \tilde{q}_\rho \tilde{q}_\nu  -\partial_\mu B_{\rho\nu} v^\rho v^\nu~,\nn
      \eea 
 where we use the notation $E^\mu_\nu = \delta^\mu_\nu + A^\mu_\nu$. One can easily check that this is a symplectomorphism 
  for any choice of tensors $A$, $B$ and $C$. If $A$ is small enough and so $E$ admits inverse $e$, $e^\mu_\nu E_\mu^\rho= \delta^\mu_\rho$ then we can easily invert these transformations and write tilde-variables in terms of those without tilde's. 
   The Hamiltonian $\Theta_{def}$ written in the deformed coordinates can be expresses as follows 
   \bea
 &&  \Theta_{def} = \tilde{p}_\mu \tilde{v}^\mu = p_\mu E^\mu_\nu v^\nu + 2 C^{\mu\nu} \tilde{q}_{\mu} \tilde{q}_\nu
    + \frac{1}{2} \Big ( E^\mu_{[\sigma} \partial_\mu E^\rho_{\nu]} + 4 C^{\rho\mu} \partial_\mu B_{\nu\sigma} \Big ) v^\nu \tilde{q}_\rho v^\sigma \nn \\
 && + E_\sigma^\mu \partial_\mu B_{\rho\nu} v^\rho v^\nu v^\sigma  
 - 2 C^{\sigma\mu} \partial_{\mu} C^{\rho\nu} \tilde{q}_\rho \tilde{q}_\nu \tilde{q}_\sigma \nn \\
 && + \Big ( C^{\rho\mu} \partial_\mu E^\sigma_\nu - C^{\sigma\mu} \partial_\mu E^\rho_\nu +
  E^\mu_\nu \partial_\mu C^{\rho\sigma} \Big ) v^\nu \tilde{q}_\rho \tilde{q}_\sigma~,
   \eea
 where 
 \be
  \tilde{q}_\mu = e_\mu^\nu q_\nu - 2 e_\mu^\nu B_{\nu\rho}v^\rho~.
 \ee
  So we can express everything in the variables without tilde's and $\Theta_{def}$ solves the master equation by definition, so it encodes the Courant algebroid. The quadratic terms encode 
   the anchor map while cubic terms encode the structure constants 
    of the Courant algebroid. If we want to set the structure constants to zero we will get 4 sets of the  equations
     which coincie exactly with the equations (\ref{eqom_MC_A-0})-(\ref{eqom_MC_A-1}) except one set 
     \be
   C^{\sigma\mu}\partial_\mu C^{\rho\nu} + {\rm cyclic~in~}(\sigma, \rho, \nu)=0~,
     \ee
which is saying that $C$ is a Poisson bi-vector. If we work in $3$-dimensions then the constraint $\partial_\mu C^{\mu\nu}=0$ (this is indeed the constraint on our fields) actually will imply that $C$ is Poisson automatically. 
Thus we see that the equations (\ref{eqom_MC_A-0})-(\ref{eqom_MC_A-1}) encode the deformations of vector fields and one forms such that the Courant brackets stay zero.

\section{A family of solutions of CME}\label{a:BRST}

We assume the algebraic data $(i-v)$ as in section \ref{s:formal}.  Let $(\M,d_L,d_R)$ be a bigraded manifold and let $(D,\D)$ be first order differential operators of bidegree $(1,0),(0,1)$, $(\gen,D^\dagger)$ second-order operators of bidegree $(0,-1),(-1,0)$ satisfying
\begin{equation}
  [D, D^\dagger]=\Box~,~~~~~~[\D, \gen] = \Box  
\end{equation}
and all other anti-commutators vanishing. As we have discussed in 
section \ref{s:formal} there is the solution for CME
\begin{equation}
    S_{BV} = \int~d\mu~ \Big ( \frac{1}{2} \A \frac{\D D}{\Box} \A + \frac{1}{6} \A^3 \Big )~,
\end{equation}
 where we assume that $\A \in {\rm Im} (\gen)$ and $\A$ is of degree $2$.  If we fix the gauge in the above action by requiring 
 $\A \in {\rm Im}(\gen) \cap {\rm Im}(D^\dagger)$
    then the gauge fixed action is invariant under two BRST symmetries
  \begin{align}
        Q \A = \Big ( D \A + \frac{1}{2} \gen (\A^2) \Big ) \Big |_{{\rm Im}(\gen) \cap {\rm Im}(D^\dagger)}~, \label{app-BRST-1}\\
        {\cal Q}\A = \Big ( \D \A - \frac{1}{2} D^\dagger (\A^2) \Big ) \Big |_{{\rm Im}(\gen) \cap {\rm Im}(D^\dagger)}~.\label{app-BRST-2}   
  \end{align}
  Since our gauge-fixed action originates from two different BV theories and the above transformations are two BV symmetries restricted to the Lagrangian submanifold then on general grounds we know that  $Q^2=0$ and ${\cal Q}^2=0$ at least on-shell. 

Let us consider the following family of first-order operators
  \begin{equation}
      \left ( \begin{array}{c}
           D'  \\
           \D'
      \end{array} \right ) = \left ( \begin{array}{cc}
      \alpha     & \beta \\
    \tilde{\alpha}    & \tilde{\beta}
      \end{array} \right) \left ( \begin{array}{c}
        D \\
        \D 
      \end{array} \right ) \equiv G \left ( \begin{array}{c}
        D \\
        \D 
      \end{array} \right )
  \end{equation}
  and second-order operators
  \begin{equation}
      \left ( \begin{array}{c}
           D'^\dagger  \\
           {\cal D}'^\dagger
      \end{array} \right ) = \left ( \begin{array}{cc}
      \gamma & \delta \\
    \tilde{\gamma}    & \tilde{\delta}
      \end{array} \right) \left ( \begin{array}{c}
        D^\dagger \\
        \gen 
      \end{array} \right )\;.
  \end{equation}
Since this family of second-order operators does not mix with the first-order ones, they still satisfy (\ref{general_requirements_2}).  
One checks that they also satisfy the same algebra (with the same $\Box$) if the following is satisfied
 \begin{equation}
     \left ( \begin{array}{cc}
      \alpha     & \beta \\
    \tilde{\alpha}    & \tilde{\beta}
      \end{array} \right)  \left ( \begin{array}{cc}
      \gamma &  \tilde{\gamma} \\
   \delta    & \tilde{\delta}
      \end{array} \right) = \left ( \begin{array}{cc}
      1     & 0\\
    0    & 1
      \end{array} \right)~,
 \end{equation}
so that $G\in GL(2)$. From now on we denote $D',\D',D'{}^\dagger,\gen{}'$ as $D_G,\D_G,D^\dagger_G,\gen_G$ respectively. By applying the construction described in \ref{s:formal}, for each $G\in GL(2)$ we can define a solution of the CME together with its gauge fixing. The gauge fixed fields are in $${\rm Im}\gen_G\cap{\rm Im}D^\dagger_G = {\rm Im} (D^\dagger) \cap {\rm Im}(\gen)~,$$
since the coefficients come from $G\in GL(2)$. The gauge-fixed action is independent of $G$ if 
   $\alpha \tilde{\beta} - \beta \tilde{\alpha}=1$ since
\begin{equation}
  D' \D' = (\alpha \tilde{\beta} - \beta \tilde{\alpha} ) D \D  ~.
\end{equation}
Thus we conclude that the gauge fixed action of each solution of the CME is independent of $G\in SL(2)$.

In general, the BV vector field is not parallel to the Lagrangian submanifold defining the gauge fixing. Nevertheless, by choosing a Lagrangian neighborhood one can project it to a vector field along the submanifold that leaves the gauge-fixed action invariant. The resulting vector field squares to zero only on shell. We call it the residual BRST symmetry (see \cite{Bonechi_2016} for details). 

We can then conclude that the gauge fixed action is invariant under the following  family of  BRST symmetries
   \begin{equation}
       Q_G \A = \Big ( (\alpha D + \beta \D) \A + \frac{1}{2} ( -\beta D^\dagger + \alpha \gen ) (\A^2) \Big ) \Big |_{{\rm Im} (D^\dagger) \cap {\rm Im}(\gen)}
   \end{equation}
   and 
   \begin{equation}
       {\cal Q}_G \A = \Big ( (\tilde{\alpha} D + \tilde{\beta}\D) \A - \frac{1}{2} (\tilde{\beta} D^\dagger -
       \tilde{\alpha} \gen) (\A^2) \Big ) \Big |_{{\rm Im} (D^\dagger) \cap {\rm Im}(\gen)}~.
   \end{equation}
  These transformations are $SL(2)$ rotations of the transformations (\ref{app-BRST-1}), (\ref{app-BRST-2}). We know
   that $Q_G = \alpha Q + \beta {\cal Q}$ and $Q_G^2=0$ on-shell,
    this implies that $Q {\cal Q} + {\cal Q} Q=0$ at least on-shell. 

Interestingly, it is possible to define two vector fields in the space of unrestricted superfields
\begin{align}
    Q \A &=  D \A + \frac{1}{2} \gen (\A^2)  - \frac{1}{2}\gen(\A)\A - \frac{1}{2}\A \gen(\A)~, \\
    {\cal Q}\A &=  \D \A - \frac{1}{2} D^\dagger (\A^2) + \frac{1}{2}D^\dagger (\A)\A+ \frac{1}{2} \A D^\dagger (\A)  \ ,
\end{align}
  which obey the same algebra as above and can be rotated by the $SL(2)$ transformations too. However, we are not aware of an action principle for which they can be interpreted as BV transformations.

\section{JT gravity and double copy}\label{ss:JT}

It is not clear to us if the double copy theory described in section \ref{s:2DYM} can 
 be identified with any known gravity theory in 2d. However, here we would like to point out the 
  structural similarities between the  double copy theory and two-dimensional Jackiw-Teitelboim (JT) gravity \cite{Teitelboim:1983ux,Jackiw:1984je} and show how the considerations from  subsection \ref{ss:so(3)}
    generalize to the 2d case. Here we work with Euclidean signature but all considerations have straightforward generalization to Minkowski case. 

 Like in the previous subsection let us consider two sets of odd coordinates: $(\theta, \zeta)= (\theta^1, \theta^2, \zeta)$ of bi-degree $(1,0)$ each and $(\theta', \zeta')=
 (\theta'^1, \theta'^2, \zeta')$ of bi-degree $(0,1)$ each. 
  We define the derivation $d=\theta^\mu\frac{\partial}{\partial x^\mu}$ and the second-order operator
  (\ref{3D-oper-SO(3)}) can be written in 2d notations as follows
 \begin{equation}
    \gen = \epsilon^{ab} \zeta' \frac{\partial^2}{\partial \theta'^a \theta'^b} + 2 \epsilon_a^{~b} \theta'^a \frac{\partial^2}{\partial \theta'^b \partial \zeta'} ~,
\end{equation}
 such that $(\gen)^2=0$ and $d\gen + \gen d=0$. We can define a  Lie bracket using $\gen$ giving rise to a dgLa, and if we restrict to ${\rm Im}(\gen)$ we still have dgLa and one can 
  define the invariant pairing 
  \begin{equation}
    \langle a, b \rangle =   \int d^2x ~d^2 \theta~d^2 \theta'~ d\zeta~d\zeta'~ ac~,~~~~b=\gen c~. 
  \end{equation}
 Thus due to the general construction in section \ref{s:formal-CS} we have a solution of the master equation 
\begin{equation}\label{JT-formal}
  S_{JT} = \frac{1}{2} \langle \A, D \A \rangle + \frac{1}{6} \langle \A, \gen (\A^2) \rangle ~,
\end{equation}
 where ${\cal A} (x, \theta, \theta', \zeta, \zeta') \in {\rm Im}(\gen)$ is superfield of bi-degree $(1,1)$ (total degree $2$) and here the BV symplectic structure is 
 \begin{equation}\label{JT-BV-str}
    \omega_{BV} = \langle \delta \A, \delta \A \rangle~.  
 \end{equation}
 Let us work out these formulas explicitly. We can describe ${\rm Im}(\gen)$ explicitly 
  as superfields linear in $\theta'$ and $\zeta'$   
\begin{equation}
  \A (x,\theta, \theta', \zeta, \zeta') = 
   \zeta \zeta' \boldsymbol{\phi} (x, \theta) + \zeta' \boldsymbol{\omega} (x, \theta) + \zeta \theta'^a \boldsymbol{\phi}_a (x,\theta)   
   + \theta'^a \boldsymbol{e}_a (x, \theta) 
\end{equation}
 such that  $\A = \gen (\tilde{\A})$ where we have 
 \begin{equation}
     \tilde{\A} = \frac{1}{4} \epsilon_{ab}
 \theta'^a \theta'^b (\zeta \boldsymbol{\phi} + \boldsymbol{\omega}) + \frac{1}{2}
  \epsilon_a^{~b} \zeta' \theta'^a (\zeta \boldsymbol{\phi}_a + \boldsymbol{e}_b)~. 
 \end{equation}
 Now working out explicitly the action (\ref{JT-formal}) we get
\begin{eqnarray}
  S_{JT} =  \int d^2x ~d^2 \theta~d^2 \theta'~ d\zeta~d\zeta'~
        \Big ( -\frac{1}{2} \A d \tilde{\A} + \frac{1}{6} \A^3 \Big )
    \end{eqnarray}
 substituting and integrating $\theta'$'s, $\zeta$ and $\zeta'$ we get
\begin{equation}\label{JT-BV-standard}
    S_{JT} = \int d^2x ~d^2 \theta~ \Big ( \boldsymbol{\phi}^a (-d \boldsymbol{e}_a + \boldsymbol{\omega} \boldsymbol{e}_a ) + \boldsymbol{\phi} ( -d \boldsymbol{\omega} + \frac{1}{2} \epsilon^{ab}
    \boldsymbol{e}_a \boldsymbol{e}_b ) \Big )
\end{equation}
 and the symplectic BV structure (\ref{JT-BV-str}) becomes 
 \begin{equation}
     \omega_{BV}= \int d^2x ~d^2 \theta~ \Big ( \delta \boldsymbol{\phi}^a \delta \boldsymbol{e}_a + \delta \boldsymbol{\phi} \delta \boldsymbol{\omega} \Big )
 \end{equation}
  where $\boldsymbol{\phi}^a (x, \theta)$, $\boldsymbol{\phi}(x,\theta)$ are
  superfields of degree $0$ and 
  $\boldsymbol{e}_a(x, \theta)$, $\boldsymbol{\omega}(x,\theta)$ are superfields of degree $1$. This is the BV formulation of 
   2d BF theory with $SO(3)$ Lie algebra, or $SO(2,1)$ for Minkowski space. Restricting the action (\ref{JT-BV-standard}) to physical field we get the standard BF formulation of JT gravity. 
    
 Although JT gravity is not equivalent to the double copy of 2d YM theory at least in any straightforward way, they share many 
  structural similarities and both theories are based on the same space of fields.

\subsection*{Competing Interests and Data Availability}
No datasets were generated or analyzed in this work. There are no
conflicts of interest applicable to this article.

 \cleardoublepage

\bibliographystyle{JHEP}
\bibliography{ref}{}

\end{document}